\documentclass[linenumbers] {aastex63}
    
    \newcommand\aastex{AAS\TeX}
     
     \usepackage{graphicx}
     \usepackage{epstopdf}
     \usepackage{longtable}
     \usepackage[utf8]{inputenc}
     \usepackage{ulem}
\received{April 22, 2021}
\revised{May 26, 2021}
\accepted{June 7, 2021}
  
\submitjournal{The Astronomical Journal}

\shorttitle{\aastex\ RR Lyrae Radial Velocities}

\shortauthors{Barnes et al.}

\begin{document}

\title{A Radial Velocity Search for Binary RR Lyrae Variables}

\correspondingauthor{Thomas G. Barnes III}
\email{tgb@astro.as.utexas.edu}

\author[0000-0002-3557-1239]{Thomas G. Barnes III}
\affil{The University of Texas at Austin, McDonald Observatory, US}

\author{Elisabeth Guggenberger}
\affil{Institute of Astronomy, University of Vienna, Austria}
\affil{Now at Hasnerstrasse 155/10, 1160 Vienna, Austria}

\author[0000-0003-0007-2939]{Katrien Kolenberg}
\affil{University of Leuven, University of Antwerp, University of Brussels, Belgium} 

\begin{abstract}
We report 272 radial velocities for 19 RR Lyrae variables. For most of the stars we have radial velocities for the complete pulsation cycle.  These data are used to determine robust center--of--mass radial velocities that have been compared to values from the literature in a search for evidence of binary systems.   Center--of--mass velocities were determined for each star using Fourier Series and Template fits to the radial velocities. Our center--of--mass velocities have uncertainties from $\pm0.16$ km s$^{-1}$ to $\pm$2.5 km s$^{-1}$, with a mean uncertainty of $\pm$0.92 km s$^{-1}$.  

We combined our center--of--mass velocities with values from the literature to look for deviations from the mean center--of--mass velocity of each star. Fifteen RR Lyrae show no evidence of binary motion (BK And, CI And, Z CVn, DM Cyg, BK Dra, RR Gem, XX Hya, SZ Leo, BX Leo, TT Lyn, CN Lyr, TU Per, U Tri, RV UMa, and AV Vir).  In most cases this conclusion is reached due to the sporadic sampling of the center--of--mass velocities over time. 

Three RR Lyrae show suspicious variation in the center--of--mass velocities that may indicate binary motion but do not prove it (SS Leo, ST Leo, and AO Peg). 

TU UMa was observed by us near a predicted periastron passage (at 0.14 in orbital phase) but the absence of additional center--of--mass velocities near periastron make the binary detection, based on radial velocities alone, uncertain.   

Two stars in our sample show $H\gamma$ emission in phases 0.9--1.0: SS Leo and TU UMa.

\end{abstract}


\section{Introduction} 

Given the number of known RR Lyrae stars, the number of confirmed RR Lyrae binaries is surprisingly low. Only two stars initially identified as RR Lyrae stars have confirmed orbits. The orbit for OGLE-BLG-RRLYR-02792 yields a mass of 0.26 \(M_\odot\), and it is now thought to be an RR Lyrae impostor (Pietrzy{\'n}ski {\em et al.} 2012). TU UMa has photometric, spectroscopic and proper motion measurements pointing consistently to binary motion with a period near 23 years (Wade {\it et al.} 1999; Li\v{s}ka {\it et al.} 2016b; Kervella {\it et al.} (2019a)). 

Numerous stars have been mentioned in the literature as possible RR Lyrae binary candidates; see the list by Li\v{s}ka {\it et al.} (2016a)\footnote{Updated on the web site \url{https://rrlyrbincan.physics.muni.cz/} by Li\v{s}ka \& Skarka.} and the review by Hajdu (2019).  Most of these candidates have been proposed based on periodic variation in the epoch of maximum light that is attributed to motion of the RR Lyrae  component about the center--of--mass of the binary. This was the case for TU UMa for which Szeidl {\it et al.} (1986) first proposed a binary explanation. However, variation in the epoch of maximum light is not {\it proof} of the binary nature.  For example, several well-determined binary orbits for RR Lyrae stars, based on variation in the epochs of maximum light, have yielded black hole candidates for the companions ({\it e.g.,} Derekas {\it et al.}  (2004),  Li\v{s}ka {\it et al.} (2016a), S\'{o}dor {\it et al.}  (2017)). None of these candidates is considered plausible by the various authors, indicating that the binary orbit interpretation is suspect. 

Kervella {\it et al.} (2019a) searched for close--in binary RR Lyrae using proper motion effects from the HIPPARCOS catalog and Gaia DR2 data.  They found significant proper motion anomalies for thirteen RR Lyrae, including only TU UMa from our current sample (discussed below).  Kervella {\it et al.} (2019b) looked for wide component RR Lyrae using the same data sets and found seven likely RR Lyrae binaries, none of which are in our sample. 

Change in the center--of--mass radial velocity (hereafter, gamma velocity) of an RR Lyrae is a clear indication of a companion.  Unfortunately the evolutionary state of RR Lyrae variables suggests that any companion will likely be of lower mass than the RR Lyrae and will likely be in a long period orbit. The radial velocity variation for the RR Lyrae component is thus likely to be small.   Moreover, if the orbit for TU UMa is typical, reflex motion of the RR Lyrae component will be particularly small for much of the orbit due to the high eccentricity.  

Fernley \& Barnes (1997) and Solano {\it et al.} (1997) have suggested RR Lyrae candidates for multiplicity based on discrepancies between the gamma velocity motions they determined and those found in the literature. For the most part, these are the stars that we have followed up in this paper. With our new data we aim to investigate the binary nature of the objects in question.  Our new, accurate, gamma velocities will also provide a foundation for multi-decade monitoring of these stars for multiplicity.  

\section{Observations}

The observations were taken at the McDonald Observatory 2.1-meter Otto Struve telescope at the f/13.5 Cassegrain focus with the Sandiford {\'E}chelle Spectrometer (McCarthy {\em et al.} 1993).  All spectra were taken at resolving power  $R=60,000$ per 2 pixels with a 1 arcsecond slit.  Most spectra covered the wavelength region 4250--4750 \AA  in 18 orders. (In one observing run the spectra were shifted 50 \AA to the blue.)  Immediately following each stellar observation and prior to a telescope move, a Th--Ar emission spectrum was obtained for wavelength calibration.  Internal bias and flat field frames were observed nightly for calibration of the Reticon 1200 x 400 CCD.  

Three observations of AV Vir were made with the 2.7m Harlan J Smith telescope at the f/33 coud{\'e} during an observing run for a different purpose (by TGB).  The Robert G Tull spectrometer (Tull  {\em et al.} 1995) was used in the ts23-TK3 mode giving $R=52,000$ for the 1.2 arcsec slit. There were 62 orders with wavelengths from 3680--10850 \AA, however only 32--33 of these orders gave useful radial velocities. These observations were reduced through wavelength calibration using IRAF\footnote{IRAF was previously distributed by the National Optical Astronomy Observatories, which are operated by the Association of Universities for Research in Astronomy, Inc., under cooperative agreement with the National Science Foundation.} by Dr. Konstanze Zwintz.  The radial velocities were determined by the authors of this paper using the same process as used for the 2.1m telescope data. 

Each night we observed the spectrum of at least one radial velocity standard star chosen from the Geneva list of CORAVEL standard stars\footnote{\url{https://obswww.unige.ch/~udry/std/}} to set the velocity zero point.  The radial velocity standard stars used for these observations are listed in Table 1. 

Basic information about each RR Lyrae is given in Table 2, primarily taken from the General Catalog of Variable Stars (Samus' {\it et al.} 2017).  Ephemerides were adopted from the sources listed.  Integration times were set not to exceed 3\% of the pulsation period to avoid smearing the velocities. The one exception is RV UMa for which the integration time was 4\% of the period due to a typo in the observing documents. 
 
\placetable{1}

\placetable{2}

\section{Data Reduction}

The CCD frames were reduced using standard IRAF procedures (Tody 1993).   The IRAF task FXCOR determines a velocity for each \'echelle order, providing multiple radial velocities for each observation. We rejected any anomalous velocities using an iterative Chauvenet's criterion (Chauvenet 1864, Taylor 1997).  To apply Chauvenet's criterion to a star we computed the mean and standard deviation from the 18 velocities in each exposure.  We determined the distance of each velocity from the mean in units of the standard deviation and computed the probability that this deviation would occur for a sample of our size.  If this probability was less than 50\%, that order was rejected.  The criterion was iterated until no rejections occurred.  The mean and standard deviation of the velocities from the remaining orders were then computed.  

We note that the number of orders retained in our reductions (varying from 11 to 17 over the sample) is a roughly linear function of the stellar metallicity and is not dependent on the S/N in the spectrum (mean S/N varying from 15 to 73; [Fe/H] values from $-$1.98 to $-$0.08). We did not make use of these relationships and only note them as information to other researchers using low S/N spectra of metal--poor stars for cross--correlation radial velocities.

The median internal uncertainty in radial velocity taken over all 272 radial velocities is $\pm 1.04$ km s$^{-1}$.  This order-to-order scatter does not include uncertainty due to any error in the adopted radial velocities of the standard stars, uncertainty due to position of the star in the spectrograph slit, uncertainty in the wavelength scale, and uncertainty due to potential differences in the velocity scale between the RR Lyrae variables and the standards used for the velocity zero point. These external uncertainties are not quantifiable from the observations. We list them here to show that the median uncertainty is solely internal and thus an underestimate of the true uncertainty. While we list the order--to--order uncertainties in Table 3, we do not use them further in this work.

\section{Radial Velocity Results}

In Table 3 we list the individual radial velocities with uncertainties, mid--point Heliocentric Julian Dates, and pulsation phases determined from the Table 2 ephemerides.

\placetable{3}

\subsection{Center--of--Mass Velocities}

For each star we computed the gamma radial velocity, first using a Fourier Series (FS) and, second, using a radial velocity template fit to the observed radial velocity curve. For the FS computation we computed the scatter in the observed radial velocities about the FS fit using only one term in the series. We added successively more terms to the series until the scatter in the fit ceased to change or until the series exhibited 'ringing' due to gaps in the phase coverage. We then adopted the zero order term in the chosen FS fit as the gamma velocity. For most stars we were limited by ringing, yielding 2-5 terms in the adopted series.  The exceptions are TT Lyn (only one observation, no FS fit), TU Per (three observations, no FS fit) and TU UMa for which we have 55 velocities and adopted a 10 term Fourier Series. 

For the template fits we distinguished between the RRab stars and the one RRc star (BX Leo). For the RRab stars we used the template published by T. Liu (1991).  For the RRc star we used the template created by Jeffery {\em et al.} (2007) based on the velocity curve of DH Peg (Jones {\em et al.} 1988). For each velocity curve we chose a phase shift between the template and the observed velocities, minimized the differences in a least--squares sense between the observed velocities and the template by shifting the template, changed the phase shift, and repeated. We chose that phase shift which minimized the velocity difference between the observations and the template. The amplitude of the T. Liu (1991) velocity template is a function of the {\em V} magnitude amplitude. We drew  {\em V} amplitudes from the literature (In order of preference: Wils {\em et al.} (2006); Automated All Sky Survey (Pojmanski (2002)), Monson  {\em et al.} (2017); and the General Catalog of Variable Stars (Samus' {\em et al.}(2017)).  For some RR Lyrae, even after scaling based on the {\em V} amplitude, the template velocity curve did not match the observed velocity curve amplitude, as would be expected for a template based on a statistical mean of velocity curves.  In these cases we arbitrarily adjusted the {\em V} magnitude amplitude until a good fit was obtained. The phase shifts ranged from $-$0.20 to 0.12, with a mean of 0.01, indicating that the ephemerides adopted are reasonable.

When our data sampled the velocity curve well, the FS method provided a much better fit to the velocity curve than did the template.  This is not surprising because the template is a statistical estimator of the individual velocity curve, whereas the Fourier Series fit approximates the velocity curve of the individual star.  When there were gaps in the phase coverage of the variable star, or the data were sparse, the template fit the velocity curve better.  When both FS and template methods fit well, we formed a weighted mean of the two gamma radial velocities. 

In Figure 1 we show both the FS fit and the Template fit to the CN Lyr data. In order to fit phases 0.8--1.0, the FS had to be of order 4.  This created ripples in the fit between phase 0.1 and 0.6. The Template was shifted $-$0.04 in phase to obtain the best fit. Both methods give the same gamma velocity: FS  23.61$\pm0.21$ km s$^{-1}$, Template 23.79$\pm0.78$ km s$^{-1}$. For this star we adopted the weighted mean of the two methods as the gamma radial velocity.

In Table 4 we list the mean HJD of our observations, the gamma radial velocity of each star, the uncertainty based on the scatter of the fit, and the type of fit (weighted mean, FS, or Template).

\placetable{4}

\section{Discussion of Individual RR Lyrae} 

In this section we compare our gamma radial velocities to historical values to determine whether evidence exists for binarity.  A search of the literature was undertaken for each of our 19 RR Lyrae. This was greatly helped by the radial velocity citations in the papers by Hemenway (1975) and Layden (1994).  For several of the historical sources, some preliminary discussion is necessary.

\subsection{Joy (1938), Joy (1950), Joy (1955)}
	Numerous gamma velocities are given in these papers, but without Julian Dates (hereafter JD) or uncertainties. For the JDs we adopted values from the papers by Abt (1970, 1973) in which he used the original observing logs by Joy to determine individual JDs for the observations. While Abt also lists the individual radial velocities, we did not recompute gamma velocities. The many decades between Joy's observations and modern ones, and the significant timespan for Joy's observations,  made it unlikely that we could phase the observations accurately. We  adopted corrections to Joy's gamma  velocities, when given, by Payne--Gaposhkin (1954). For uncertainties we estimated a typical value by comparing the gamma velocities given by Joy, as corrected, with all other values from the literature for each star.  This yielded an rms uncertainty of $\pm 28$ km s$^{-1}$ for nine of the thirteen stars in common with us. The other four stars have very large differences between Joy's values and recent values and were ignored in forming the mean uncertainty. We adopted $\pm 28$ km s$^{-1}$ for all gamma velocities determined by Joy.
	
\subsection{Layden (1994)}
	This publication gives gamma velocities with uncertainties for seventeen of our targets.  Four of these have very large residuals compared to other sources.  The remaining thirteen stars give an average rms uncertainty of $\pm 26$ km s$^{-1}$, which compares well with the typical uncertainty quoted by Layden (1994): $\pm 2$ km s$^{-1}$ to $\pm 30$ km s$^{-1}$. We adopted the individual uncertainties given by Layden.  In a few cases, Layden gave radial velocities but not gamma velocities. In these cases we fit templates to his velocities as noted below for the pertinent stars. 
	
\subsection{S.  Liu {\em et al.} (2013)}
	This publication provided five radial velocities for four stars in our sample, but did not compute gamma velocities or publish velocity uncertainties.  As HJDs were given, we fit templates to the radial velocities of the four stars, using the ephemerides of Table 2, to determine gamma velocities.  The rms deviations of these four gamma velocities from those in the literature is  $\pm 5$ km s$^{-1}$, which we adopted as the uncertainty for each gamma value. 
	
\subsection{G.--C. Liu {\em et al.} (2020)}
	Eight stars in our sample have gamma radial velocities with uncertainties in the catalog by G.--C. Liu {\em et al.} (2020).  One star has a very large residual compared to our sample.  The remaining seven stars give an rms deviation from our sample of $\pm 21$ km s$^{-1}$ which is within the typical uncertainties quoted by G.--C. Liu {\em et al.} (2020): $\pm 5$ km s$^{-1}$ to $\pm 21$ km s$^{-1}$.  Unfortunately no JDs are published for these eight stars. In order to compare their gamma radial velocities with our sample, we have estimated a common JD for these stars as the mid-point of the observing sample listed by G.--C. Liu {\em et al.} (2020).  {\em We emphasize that this is done for illustrative purposes only as the true JD for each gamma velocity is unpublished.}  We infer that the spectra used by G.--C. Liu {\em et al.} (2020) were observed in the interval 2006--2013, from which we have chosen a mid-point JD of 2455197. When comparing the G.-C. Liu {\em et al.} (2020) gamma radial velocities with the rest of our sample in the discussions below, bear in mind that this illustrative JD may be in error by {$\pm 4$ years.
	
\subsection{Comments on individual stars}

	Table 5 contains the gamma velocities found in the literature and our results. We list for each source the mean JD of the observations, the gamma velocity and its uncertainty, the residual from the weighted mean of all gamma velocities for that star, and the source.  When the JD has been estimated, ({\em e. g.,} see above for  G.--C. Liu {\em et al.} (2020)), we have put the entire entry in the table in italics. 
	
\subsubsection{BK And}	
	This star was added to our observing list because of disagreement in the gamma results by Layden (1994) and Jeffery {\em et al.} (2007).  There are three gamma radial velocities spanning 25 years in the literature which are given in Table 5 and shown in Figure 2. Layden (1994) published a single radial velocity for this star but did not publish a gamma value. We fit a template to his observation, using the ephemerides of Table 2, to obtain the gamma value in Table 5. We adopted Layden's uncertainty on the single observation as the gamma uncertainty.  The weighted mean of the three gamma values is $-$9.47 km s$^{-1}$ with an uncertainty of $\pm 0.25$ km s$^{-1}$.
	
	BK And is not in the list by Li\v{s}ka {\it et al.} (2016a) as a suspected binary. We see no evidence of binary motion for this star.

\subsubsection{CI And}	
	This star was added to our observing list because of disagreement in the gamma results of Layden (1994) and Fernley \& Barnes (1997) .  There are five gamma radial velocities spanning 25 years in the literature which are given in Table 5 and shown in Figure 3.  S. Liu {\em et al.} (2013) published two radial velocities for this star but did not publish a gamma value. We fit a template to these observations, using the ephemerides of Table 2, to obtain the gamma values and uncertainties in Table 5. As noted above, the JD for the gamma velocity by G.-C. Liu {\em et al.} (2020) is illustrative only.  The weighted mean of the five gamma values is 19.74 km s$^{-1}$ with an uncertainty of $\pm 0.88$ km s$^{-1}$.  
	
	CI And is in the list by Li\v{s}ka {\it et al.} (2016a) as a suspected binary.  Fernley \& Barnes (1997) noted a 76 km s$^{-1}$ difference between their gamma velocity and that by Layden (1994) and suggested possible binary motion.  The joint uncertainty of the Layden (1994) and Fernley \& Barnes (1997) gamma values is 33 km s$^{-1}$, indicating a $2.30\sigma$ difference.  Our new gamma value supports the result of Fernley \& Barnes (1997). 
	
	We see no evidence of binary motion for this star.
	
\subsubsection{Z CVn}
	
	This star was added to our observing list because of variations in the time of maximum light.  There are six gamma radial velocities spanning 92 years in the literature which are given in Table 5 and shown in Figure 4.  Skarka {\it et al.} (2018) reanalyzed the radial velocity measures by Joy (1938); we adopted their new gamma value and uncertainty. As noted above the JD for the gamma velocity by G.--C. Liu {\em et al.} (2020) is illustrative only. The weighted mean of the six gamma values is $-$0.30 km s$^{-1}$ with an uncertainty of $\pm 0.66$ km s$^{-1}$. 

	Z CVn is in the list by Li\v{s}ka {\it et al.} (2016a) as a suspected binary.  Skarka {\it et al.} (2018) performed an extensive analysis of the variation in times of maximum light. The timing variations suggest an orbit with a 78.3 year period and an eccentricity of 0.63.  However, their analysis of the historical gamma velocities shows no evidence for such an orbit. (We note that preliminary radial velocities from the current work were used in their analysis.  Those values are supplanted and extended by the values in Table 3.  The conclusions of Skarka {\it et al.} (2018) are not affected by these minor changes.)  The binary orbit predicted periastron passages on JD2427550 and JD2456149, which are in our radial velocity window, with a periastron radial velocity of order 57 km s$^{-1}$.  As can be seen in Figure 4, the center--of--mass radial velocities do not support the binary orbit, in agreement with Skarka {\it et al.} (2018).  Moreover, their orbit suggests a minimum mass for the unseen companion to the RR Lyrae of 56.5 $M_\sun$.  The only plausible candidate with this mass would be a black hole, which the authors reject. They reject the binary hypothesis for Z CVn.
	
	Given the very large scatter in the gamma values, we see no evidence of binary motion.  
	
\subsubsection{DM Cyg}	
	This star was added to our observing list because of disagreement in the gamma results of Layden (1994) and Fernley \& Barnes (1997).  There are six gamma radial velocities spanning 76 years in the literature which are given in Table 5 and shown in Figure 5. S. Liu {\em et al.} (2013) published one radial velocity for this star but did not publish a gamma value. We fit a template to this observation, using the ephemerides of Table 2, to obtain the gamma value in Table 5.  The weighted mean of the six gamma values is $-$21.79 km s$^{-1}$ with an uncertainty of $\pm 0.74$ km s$^{-1}$.

	DM Cyg is in the list by Li\v{s}ka {\it et al.} (2016a) as a suspected binary.  Fernley \& Barnes (1997) noted a 94 km s$^{-1}$ difference between their gamma velocity and that by Layden (1994) and suggested possible binary motion.  The joint uncertainty of the Layden (1994) and Fernley \& Barnes (1997) gamma values is 33 km s$^{-1}$, indicating a $2.85\sigma$ difference. Our new gamma value supports the result of Fernley \& Barnes (1997). 

	Four of the six gamma values are concordant, with the other two discrepant but with large uncertainties. We see no evidence of binary motion for this star.
	
\subsubsection{BK Dra}	
	This star was added to our observing list because of disagreement in the gamma results by Layden (1994) and Fernley \& Barnes (1997).  There are three gamma radial velocities spanning 27 years in the literature which are given in Table 5 and shown in Figure 6. The weighted mean of the three gamma values is $-$85.22 km s$^{-1}$ with an uncertainty of $\pm 1.11$ km s$^{-1}$.
	
	We have one anomalous radial velocity for this star.  On HJD2457562.85 we obtained a radial velocity of $-$23.02 km s$^{-1}$ with an internal uncertainty of $\pm 1.45$ km s$^{-1}$. This value is 43 km s$^{-1}$ more positive than our well-defined radial velocity curve at that phase (0.73). Despite a thorough examination of the data and reductions, we have not found a reason for this discrepancy.  While we suspect that there is an undetected observational error in this value, we report it for completeness. This datum was excluded from our fit for the gamma radial velocity. 
	
	BK Dra is in the list by Li\v{s}ka {\it et al.} (2016a) as a suspected binary.  Fernley \& Barnes (1997) noted a 54 km s$^{-1}$ difference between their gamma velocity and that by Layden (1994) and suggested possible binary motion.  However, the joint uncertainty of the Layden (1994) and Fernley \& Barnes (1997) gamma values is 40 km s$^{-1}$, indicating only a $1.35\sigma$ difference. We do not judge the difference to be significant. Our new gamma value supports the result of Fernley \& Barnes (1997). 
	
	Li {\it et al.}  (2018) analyzed variations in the time of maximum light over 118 years and proposed a binary period of 78.4 years with an eccentricity of 0.70 (their Model 2). In their model, periastron passage was at JD2443717 $\pm378$ days.  As can be seen in Figure 6, this periastron passage is well before the first gamma value we have (10 years prior).  The next periastron passage is predicted to be JD2472353 (late 2056). 
	
	Given the predictions of a highly eccentric orbit and a long period, we cannot confirm the binary nature of BK Dra.  The gamma value given here may be used in the future as a baseline for possible radial velocity motion near the next predicted periastron. 
	
\subsubsection{RR Gem}	
	This star was added to our observing list because of the suspected binary motion noted by Firmanyuk (1976).  There are six gamma velocities spanning 90 years in the literature which are given in Table 5 and shown in Figure 7.   As noted above the JD for the gamma velocity by G.--C. Liu {\em et al.} (2020) is illustrative only.  The weighted mean of the six gamma values is 65.57 km s$^{-1}$ with an uncertainty of $\pm 0.72$ km s$^{-1}$.
	
	RR Gem is in the list by Li\v{s}ka {\it et al.} (2016a) as a suspected binary.  Firmanyuk (1976) analyzed variations in the time of maximum light over 68 years to estimate an orbital period of 25600 days ($\sim$70 years). He did not publish a date for the hypothetical periastron passage and its radial velocity.  However, he noted that the results are unsatisfactory as the sum of the masses for the suspected binary is in the range 100--1000 $M_{\sun}$.  
	
	With the exception of the very uncertain gamma velocity by Joy (1938) and the gamma velocity by G.--C. Liu {\it et al.} (2020), the measures cluster reasonably about the mean value. We see no evidence of binary motion for this star.
	
\subsubsection{XX Hya}

	This star was added to our observing list because of disagreement in the gamma results of Fernley \& Barnes (1997) and Solano {\it et al.} (1997).  There are five gamma radial velocities spanning 72 years in the literature which are given in Table 5 and shown in Figure 8.  The weighted mean of the five gamma values is 65.97 km s$^{-1}$ with an uncertainty of $\pm 1.10$ km s$^{-1}$. 
	
	Abt (1970) did not list a JD for the single observation going into Joy's (1950) published gamma velocity.  We made a rough estimate for the probable observation date near JD2431091. {\em This is done for illustrative purposes only and may be in error by as much as six years.}  

	XX Hya is in the list by Li\v{s}ka {\it et al.} (2016a) as a suspected binary based on the disagreement noted above. There is a 63 km s$^{-1}$ difference between Fernley \& Barnes (1997) and Solano {\it et al.} (1997).  The joint uncertainty of the two gamma values is 40 km s$^{-1}$, 1.6$\sigma$.  We do not judge the difference to be significant. 
	
	Given the very large scatter, and large uncertainties, we see no evidence of binary motion. 

\subsubsection{SS Leo}

	This star was added to our observing list because it is well-known to show variation in the time of maximum light.  There are nine gamma velocities spanning 81 years in the literature which are given in Table 5 and shown in Figure 9. Fernley {\it et al.} (1990) published 63 radial velocities for this star but did not publish a gamma value. We fit those data with both Fourier Series and a Template, using the ephemerides of Table 2, to obtain gamma values and uncertainties.  Both methods agreed on the gamma value; the template fit had a much better uncertainty so that result is included in Table 5. As noted above the JD for the gamma velocity by G.-C. Liu {\em et al.} (2020) is illustrative only. The weighted mean of the nine gamma values is 161.74 km s$^{-1}$ with an uncertainty of $\pm 0.28$ km s$^{-1}$. 
	
	SS Leo is in the list by Li\v{s}ka {\it et al.} (2016a) as a suspected binary.  Li\v{s}ka {\it et al.} (2016a) analyzed variations in the time of maximum light over 118 years to propose an orbital period of 110.7 years with an uncertainty near 6 years. According to their Table 3, periastron passage occurred at JD2453240 with an uncertainty near 500 days, which is the only proposed periastron passage in our radial velocity window.  Li\v{s}ka {\it et al.} (2016a) found an orbital $K_1$  value for SS Leo of 2.73 km s$^{-1}$. Three gamma values (excluding the value of G.-C. Liu {\em et al.} (2020)) starting 1711 days (0.04 in orbital phase) after predicted periastron passage are suggestive of a gamma deviation $\ge$3 km s$^{-1}$. 	
	
	The uncertainties are too large to definitely confirm binary motion, but the gamma results are certainly suggestive.
	
\subsubsection{ST Leo}
	This star was added to our observing list because Fernley \& Barnes (1997) noted disagreement among the gamma values of  previous works and theirs. There are seven gamma velocities spanning 82 years in the literature which are given in Table 5 and shown in Figure 10.  Rogers (1960) and Woolley \& Savage (1971) published gamma values for ST Leo, but no JDs are given. In order to compare their gamma radial velocities in a time sequence with our sample, we have estimated the JDs: 2436204.5 for Rogers (1960), two years prior to publication, and 2438799.5 for Woolley {\em et al.} (1965), six years prior to publication. The latter date is highly uncertain as it is based on unpublished observations by A. Sandage.  {\em We emphasize that this is done for illustrative purposes only as the true JDs are unpublished.}  When comparing these two gamma radial velocities with the rest of our sample in the discussion below, bear in mind that the illustrative JDs may be in error by several years.  The weighted mean of the seven gamma values is 166.08 km s$^{-1}$ with an uncertainty of $\pm 1.13$ km s$^{-1}$. 
	
	ST Leo is in the list by Li\v{s}ka {\it et al.} (2016a) as a suspected binary because of the comments by Fernley \& Barnes (1997). The scatter in the historical gamma values is indeed larger than would be expected on the basis of their uncertainties; {\em e.g.,} compare the values from Rogers (1960), Fernley \& Barnes (1997), Chadid {\it et al.} (2017) and this work.  The quoted uncertainties are much smaller than the scatter. This could be a result of underestimated uncertainties rather than binary motion.
		
	The uncertainties are too large to confirm binary motion, but the gamma velocity scatter is suggestive.
	
\subsubsection{SZ Leo}
	This star was added to our observing list because of the large differences between the gamma values of Joy (1938), Hawley \& Barnes (1985) and Layden (1994). There are five gamma velocities spanning 83 years in the literature, which are given in Table 5 and shown in Figure 11. As noted above the JD for the gamma velocity by G.-C. Liu {\em et al.} (2020) is illustrative only. The weighted mean of the five gamma values is 171.01 km s$^{-1}$ with an uncertainty of $\pm 0.73$ km s$^{-1}$. 
	
	SZ Leo is not in the list by Li\v{s}ka {\it et al.} (2016a) as a suspected binary. The scatter in the historical gamma values is larger than would be expected on the basis of their uncertainties; {\em e.g.,} compare the values from Hawley \& Barnes (1985), G.--C. Liu (2020), and this work.  

	The uncertainties are too large to confirm binary motion.
	
\subsubsection{BX Leo}	
	This RRc star was added to our observing list because of disagreement in the gamma results of Fernley \& Barnes (1997) and Solano {\it et al.} (1997).  There are three gamma radial velocities spanning 21 years in the literature which are given in Table 5 and shown in Figure 12.  The weighted mean of the three gamma values is 10.35 km s$^{-1}$ with an uncertainty of $\pm 0.77$ km s$^{-1}$.
	
	Both the Fourier Series and Template fits to the velocity curve of BX Leo show large scatter, $\sim$3 times the internal uncertainties in the individual radial velocities.  Monson {\it et al.} (2017) noted that, for a non-Blazhko variable, BX Leo was particularly difficult to phase in the light curve.  A similar comment was made by Neeley  {\it et al.} (2019) in attempting to phase the Gaia DR2 photometry. While our radial velocity observations only cover a time interval of 212 days, if the scatter in them is due to a close companion, we should see evidence of that in the residuals. We looked for such evidence by examining both the Fourier Series residuals and the Template residuals as a function of time.  In both cases the residuals show no trend. This suggests an origin of the phasing problem in BX Leo itself.  We suggest that this star would be a good candidate for an intense photometric campaign. 

	BX Leo is in the list by Li\v{s}ka {\it et al.} (2016a) as a suspected binary based on the gamma discrepancy noted above.  The difference between the Fernley \& Barnes (1997) and Solano {\it et al.} (1997) gamma values is 37 km s$^{-1}$ with a  joint uncertainty of 18 km s$^{-1}$, indicating a $1.88\sigma$ difference.  We judge this to be insignificant. Our new gamma value lies between these two values. 
	
	Given the uncertainties we see no evidence of binary motion for this star.

\subsubsection{TT Lyn}
	We have a single observation of TT Lyn, inadvertently made as part of this program, that we include here. There are eight gamma radial velocities spanning 49 years in the literature, which are given in Table 5 and shown in Figure 13. Layden (1994) published two radial velocities of TT Lyn but did not compute a gamma value. We fit a template to his observations, using the ephemerides of Table 2, to obtain a gamma value and uncertainty. The gamma velocity JD by G.--C. Liu {\em et al.} (2020) is illustrative only. The weighted mean of the eight gamma values is $-$62.87 km s$^{-1}$ with an uncertainty of $\pm 1.09$ km s$^{-1}$. 

	TT Lyn is not in the list by Li\v{s}ka {\it et al.} (2016a) as a suspected binary. The scatter in the historical gamma values is consistent with the uncertainties except for two values.  Compared to their uncertainties, the Layden (1994) value is too high by 12 km s$^{-1}$, and that by G.--C. Liu {\it et al.} (2020) is too low by $-$24 km s$^{-1}$s. Quite likely our template fit to the two Layden (1994) radial velocities gave too small an uncertainty as the individual radial velocities have uncertainties of 22 km s$^{-1}$ and 29 km s$^{-1}$.
	
	The gamma velocity measures cluster reasonably about the mean value, with those two exceptions. We see no evidence of binary motion for this star.
	
\subsubsection{CN Lyr}	
	This star was added to our observing list because of disagreement in the gamma results of Layden (1994) and Fernley \& Barnes (1997). There are five gamma radial velocities spanning 24 years in the literature which are given in Table 5 and shown in Figure 14. S. Liu {\em et al.} (2013) published two radial velocities for this star but did not publish a gamma value. We fit a template to these observations, using the ephemerides of Table 2, to obtain the gamma value and uncertainty in Table 5.  The weighted mean of the five gamma values is 23.64 km s$^{-1}$ with an uncertainty of $\pm 0.21$ km s$^{-1}$.
	
	CN Lyr is in the list by Li\v{s}ka {\it et al.} (2016a) as a suspected binary.  Fernley \& Barnes (1997) noted a 40 km s$^{-1}$ difference between their gamma velocity and that by Layden (1994) and suggested possible binary motion.  The joint uncertainty of the Layden (1994) and Fernley \& Barnes (1997) gamma values is 33 km s$^{-1}$, indicating a $1.21\sigma$ difference.  We judge this difference to be insignificant. Our new gamma value supports the result of Fernley \& Barnes (1997). 
	
	We see no evidence of binary motion for this star.

\subsubsection{AO Peg}
	This star was added to our observing list because of disagreement among the gamma results by Joy (1950), Layden (1994) and Jeffery {\em et al.} (2007). There are six gamma radial velocities spanning 67 years in the literature which are given in Table 5 and shown in Figure 15.  Layden (1994) and S. Liu {\em et al.} (2013) each published two radial velocities for this star but did not publish gamma values. We fit templates to these observations, using the ephemerides of Table 2, to obtain the gamma values and uncertainties in Table 5. The JD for the gamma velocity by G.--C. Liu {\em et al.} (2020) is illustrative only. The weighted mean of the six gamma values is $-$269.25 km s$^{-1}$ with an uncertainty of $\pm 0.11$ km s$^{-1}$. 
	
	AO Peg is not in the list by Li\v{s}ka {\it et al.} (2016a) as a suspected binary. The deviations of the Layden (1994) and G.-C. Liu {\em et al.} (2020) values from the other four are suggestive but not conclusive of binary motion.
	
\subsubsection{TU Per}
	This star was added to our observing list because of the large differences between the gamma values of Joy (1938) and Jeffery {\em et al.}  (2007). There are five gamma radial velocities spanning 91 years in the literature, which are given in Table 5 and shown in Figure 16.  Hawley \& Barnes (1985) published one radial velocity of TU Per to which we fit a template, using the ephemerides of Table 2, to obtain the gamma velocity given in Table 5.  For its uncertainty we estimated $\pm 10$ km s$^{-1}$ based on the uncertainty quoted by Hawley \& Barnes (1985). The JD for the gamma velocity by G.-C. Liu {\em et al.} (2020) is illustrative only. The weighted mean of the five gamma values is $-$313.86 km s$^{-1}$ with an uncertainty of $\pm 0.52$ km s$^{-1}$. 
	
	TU Per is not in the list by Li\v{s}ka {\it et al.} (2016a) as a suspected binary.  At face value the distribution of gamma values suggests a long--term linear trend.  However, the gamma value published by Joy (1938), while far from the mean of the other four values, is only $1.8\sigma$ deviant, which we judge to be insignificant. 
	
	The gamma measures cluster reasonably about the mean value.   We see no evidence of binary motion, but we suggest future monitoring of TU Per in gamma velocity to determine the reality (or not) of the long--term trend.
		
\subsubsection{U Tri}
	This star was added to our observing list because of disagreement in the gamma results of Layden (1994) and Jeffery {\em et al.} (2007).  There are four gamma radial velocities spanning 76 years in the literature which are given in Table 5 and shown in Figure 17. The weighted mean of the four gamma values is $-$37.86 km s$^{-1}$ with an uncertainty of $\pm 0.76$ km s$^{-1}$.
	
	U Tri is not in the list by Li\v{s}ka {\it et al.} (2016a) as a suspected binary. We note a 93 km s$^{-1}$ difference between Layden (1994) and Jeffery {\em et al.} (2007).  The joint uncertainty of the Layden (1994) and Jeffery {\em et al.} (2007) gamma values is 31 km s$^{-1}$, indicating a $3\sigma$ difference. Our new gamma value supports the result of Jeffery {\em et al.} (2007).
	
	Given the paucity of gamma velocities and the large uncertainties of the pre-1995 measures, we see no evidence of binary motion.
	
\subsubsection{RV UMa}
	This star was added to our observing list because it is well-known to show variation in the time of maximum light.  There are six gamma radial velocities spanning 95 years in the literature which are given in Table 5 and shown in Figure 18. Fernley {\em et al.} (1993) published five radial velocities for this star but did not publish a gamma value. We fit a template to these observations, using the ephemerides of Table 2, to obtain the gamma value and uncertainty in Table 5.  The weighted mean of the six gamma values is $-$187.67 km s$^{-1}$ with an uncertainty of $\pm 0.43$ km s$^{-1}$. 
	
	RV UMa is in the list by Li\v{s}ka {\it et al.} (2016a) as a suspected binary.  Li\v{s}ka {\it et al.} (2016a) analyzed variations in the time of maximum light over 115 years to propose an orbital period of 66.9 years. According to their Table 3, periastron passage occurred at JD2449520 with an uncertainty near 550 days, which is in our radial velocity window as is the previous periastron passage.  There are two gamma values approximately 0.06 in orbital phase preceding the later predicted periastron passage; those by Layden (1994) and Fernley {\it et al.} (1993). The Layden value is too uncertain to be useful in this context, and the Fernley {\it et al.} value, while having a very low uncertainty, is sensibly the same as our new value, also with a very low uncertainty, obtained at orbital phase 0.32. Li\v{s}ka {\it et al.} (2016a) determined an orbital $K_1$  value for RV UMa of 2.24 km s$^{-1}$, which we should have seen given the uncertainties of the two gamma values. 

	The gamma measures cluster reasonably about the mean value. Given the uncertainties we do not detect evidence of binary motion for this star.
	
\subsubsection{TU UMa}
	This star was added to our observing list because it is well--known as a probable binary. There are eleven gamma velocities spanning 84 years in the literature, which are given in Table 5 and shown in Figure 19.  Preston \& Paczynski (1964) published twelve radial velocities to which we fit a template, using the ephemerides of Table 2, to obtain the gamma velocity and uncertainty given in Table 5. Britavskiy {\it et al.} (2018) computed gamma velocities using two methods; we have taken the mean of those two values for Table 5. The weighted mean of the eleven gamma values is 93.75 km s$^{-1}$ with an uncertainty of $\pm 0.16$ km s$^{-1}$. 
	
	TU UMa is in the list by Li\v{s}ka {\it et al.} (2016a) as a suspected binary.  Li\v{s}ka {\it et al.} (2016b) analyzed variations in the time of maximum light over $\sim$100 years to propose an orbital period of 23.30 years with an uncertainty of 0.06 years. According to their Table 3, Model 1, periastron passage occurred at JD2447092 with an uncertainty near 40 days, which is in our radial velocity window, as are several adjacent periastron passages. These are marked in Figure 19 at the predicted minimum radial velocity. 
		
	The Li\v{s}ka {\it et al.} (2016b) analysis contained the same set of gamma velocities as in our Table 5, except for the recent values from Britavskiy {\it et al.} (2018) and ourselves. While the scatter in the gamma velocities is large (only seven of their fifteen gamma velocities have one sigma uncertainties that lie on the predicted orbit), they concluded that the radial velocity observations support the binary orbit.  The new results by Britavskiy {\it et al.} (2018) and ourselves are at orbital phases 0.92 and 0.14, respectively. At these phases the binary orbit predicts gamma velocities $\sim$89 km s$^{-1}$ and $\sim$92 km s$^{-1}$, respectively.  The Britavskiy {\it et al.} (2018) gamma velocity is 14 km s$^{-1}$ $\pm 2$ km s$^{-1}$ high and our value is 2 km s$^{-1}$ $\pm 0.1$ km s$^{-1}$ high, compared to the predicted orbit.  These new gamma velocities may help improve the orbit. The binary orbit predicts measurable reflex motion during a periastron passage that is only $\sim$0.2 in orbital phase wide (1703 days) which makes confirmation of periastron passage difficult by historical radial velocity measures. The next opportunity for a concerted radial velocity campaign on TU UMa does not occur until mid--2034.
	
	Kervella {\em et al.} (2019a) repeated the Li\v{s}ka {\it et al.} (2016b) analysis, adding proper motion anomalies from the HIPPARCOS catalog and Gaia DR2 data.  They determined several orbital parameters additional to the Li\v{s}ka {\it et al.} (2016b) parameters and concluded that the probable mass of the unseen component is 1.98$\pm0.33$ $M_{\sun}$.  As they note, the mass is high for an unseen component, implying either a massive white dwarf or a neutron star. 
	
	The gamma velocity measures in Figure 19 are suggestive of binary motion but do not confirm it in our opinion.  However, the strengths of the Li\v{s}ka {\it et al.} (2016b) and Kervella {\em et al.} (2019a) analyses are persuasive. 
	
\subsubsection{AV Vir}
	 This star was added to our observing list because of the large differences among the gamma results by Joy (1950), Hawley \& Barnes (1985) and Layden (1994). There are four gamma radial velocities spanning 67 years in the literature which are given in Table 5 and shown in Figure 20. The weighted mean of the four gamma values is $-$151.43 km s$^{-1}$ with an uncertainty of $\pm 0.65$ km s$^{-1}$.  
	
	AV Vir is not in the list by Li\v{s}ka {\it et al.} (2016a) as a suspected binary. We see no evidence of binary motion for this star.
	
\subsubsection{$H\gamma$ emission}
Some RR Lyrae variables are known to show hydrogen in emission during the rising branch of the light curve.  This is attributed to propagation of a shock front through the atmosphere: Gillet \& Crowe (1988), Chadid {\em et al.} (2008), Gillet {\em et al.} (2019).

We examined all of our spectra that were taken in the phase interval 0.80--1.00 for $H\gamma$ emission. BK And, TT Lyn, and TU Per were not observed at these phases.  Only SS Leo and TU UMa showed evidence of emission. 

SS Leo showed blue--shifted $H\gamma$ emission at phase 0.94; the only observation in the phase window. 

TU UMa showed strong blue--shifted $H\gamma$ emission in four spectra at phases (0.960), (0.933, 0.997), and (0.932). (The parentheses group phases in the same pulsation cycle together.) Very weak $H\gamma$ emission was possibly present at phases 0.000 (the same cycle as phase 0.960 above) and 0.037 (the same cycle as 0.933 and 0.997 above).  No $H\gamma$ emission was seen in two spectra at phases 0.893 (the same cycle as 0.933 and 0.997) and 0.995 (a fourth cycle). The presence of emission at phase 0.997 and its absence at phase 0.995 in a different pulsation cycle indicates that the shock behavior changes from cycle to cycle. In Figure 21 we show the emission feature for the spectrum taken at phase 0.933. 

In visual inspection we did not see line--doubling in SS Leo and TU UMa near the phases of $H\gamma$ emission.  Likewise, none of the FXCOR cross-correlation functions at phases near the emission showed evidence of line-doubling.

Extending our analysis to $H\gamma$  emission is well beyond the intent of this paper.  We comment that we did not see it in most of our sample and did see it in two stars as an aid to other researchers. Any researcher wishing to use these spectra in an analysis is welcome to contact us for them.

\section{Conclusions}
We have presented 272 new radial velocities for a set of 19 RR Lyrae suspected of being members of binaries. Most of our radial velocity curves cover complete pulsation cycles.  Center--of--mass velocities were determined for each star using Fourier Series and Template fits to the velocities. The new center--of--mass velocities have uncertainties from $\pm0.16$ km s$^{-1}$ to $\pm$2.5 km s$^{-1}$, with a mean uncertainty of $\pm$0.92 km s$^{-1}$.  These uncertainties are sufficient to detect binary motion should the observations be taken near periastron passage in an elliptical orbit.  

We combined our new center--of--mass velocities with values from the literature to look for deviations from the mean center--of--mass velocity of each star. Fifteen RR Lyrae show no evidence of binary motion (BK And, CI And, Z CVn, DM Cyg, BK Dra, RR Gem, XX Hya, SZ Leo, BX Leo, TT Lyn, CN Lyr, TU Per, U Tri, RV UMa, and AV Vir).  In most cases this conclusion is reached due to the sporadic sampling of the center--of--mass velocities over time. Nonetheless we are able to reject most assertions in the literature that large differences in historical gamma velocities for these stars suggest binary motion. 

Three RR Lyrae show suspicious variation in the center--of--mass velocities that may indicate binary motion but do not prove it (SS Leo, ST Leo, and AO Peg). These stars are worthy of additional radial velocity measures. 

TU UMa was observed by us near a predicted periastron passage (at 0.14 in orbital phase) but the absence of additional center--of--mass velocities near periastron make the binary detection, based on radial velocities, uncertain.  Other analyses using periodic variation in the epoch of maximum light and proper motion anomalies strongly suggest binary motion. There is a predicted periastron passage for TU UMa  in mid-2034 that should be the subject of a concerted radial velocity campaign.  

Two stars showed H$\gamma$ emission in the phase window 0.9--1.0 as discussed above: SS Leo and TU UMa.

The RRc star BX Leo showed excessive scatter in its radial velocity curve, which was also noted by other authors for the photometric light curves. We searched for evidence of radial velocity variation due to a short period binary in our 212 day observing window and found none. We suggest that the excessive scatter is intrinsic to the star. 

\acknowledgments
We thank Dr. Konstanze Zwintz for reduction of the 2.7m telescope observations of AV Vir. Drs. George Preston and Chris Sneden provided unpublished JDs for SS Leo and ST Leo. TGB thanks Dr. John Kuehne for help in getting important software operating on his computer.  Many thanks to Ms. Lydia Fletcher for copying two, critical research papers in the UT Austin libraries during Covid19 closure of the university.  We thank an anonymous referee for recommendations that improved this paper.

\vspace{5mm}
\facilities{Struve, Smith}

\newpage

 \begin{deluxetable}{rccclc}
\tablenum{1}
\tablewidth{0pt}
\tablecaption{Adopted Radial Velocity Standard Stars}
\tablehead{\colhead{Star} &\colhead{RA} &\colhead{Dec} &\colhead{V}
&\colhead{Spectral} &\colhead{Radial Velocity}\\&(2000) &(2000) & (mag) & Class &(km s$^{-1}$)}
\startdata
HD22879 & 03:40:22 & -03:13:01 & 6.68 & F9 V & 120.2 \\
HD65934 & 08:02:11 &  +26:38:16 &  7.70 &  G8 III & 35.8  \\
HD102870 & 11:50:42 & +01:45:53 & 3.59 & F8 V  & 4.3  \\
HD136202 & 15:19:19 & +01:45:55 & 5.04 & F8 III--IV  &  54.3 \\
HD154417 & 17:05:17 & +00:42:09 & 6.00 & F9 V & -16.8 \\
HD187691 & 19:51:02 & +10:24:57 &5.12 & F8 V & 0.0 \\
HD222368 & 23:39:57 & +05:37:35 & 4.13 & F7 V  &  5.6 \\
\enddata
\end{deluxetable}    
                                                                         
\begin{deluxetable}{rccclllc}
\tablenum{2}
\tablewidth{0pt}
\tablecaption{Observed RR Lyrae Stars}
\tablehead{\colhead{Star} &\colhead{RA} &\colhead{Dec} &\colhead{V mag} & \colhead{HJD-2400000} & Period 
 & \colhead{Source} &\colhead{Integration Time}\\ &(2000) &(2000) & (mag) & (max light)  & (days) & & (seconds)}
\startdata
CI And &  01:55:08.3 &  +43:45:56 &  11.76--12.66  & 52695.3060 & 0.484728 & Maintz (2005) & 1250 \\
U Tri &  01:55:31.4 & +33:46:08 & 11.88--13.20  & 51952.2760 & 0.447253& Maintz (2005) & 1150 \\
TU Per & 03:09:04.8& +53:11:36 & 11.94--13.05   & 52991.3060 &0.60707490 & Maintz (2005) &1550 \\
RR Gem & 07:21:33.5 & +30:52:59 & 10.62--11.99 &56750.485  &0.39729 &Monson {\em et al.} (2017) & 1020 \\
TT Lyn & 09:03:07.8 & +44:35:08 & 9.42--10.21 &  56750.790&0.597434355 & Monson  {\em et al.} (2017) &1500 \\
XX Hya & 09:09:49.5 & $-$15:35:59 & 11.24--12.49  &39832.0110  &0.50776718 &Maintz (2005)  &1300  \\
SZ Leo &11:01:36.8 & +08:09:56 & 11.91--12.79 & 53107.4110 &0.53406125 & Maintz (2005)&1380  \\
TU UMa & 11:29:48.5 & +30:04:02 & 9.26--10.24 & 56570.033 &0.5576587 &Monson  {\em et al.} (2017)  &1445  \\        
SS Leo & 11:33:54.5 & $-$00:02:00 &10.38--11.56 & 53050.5650&0.626335 & Maintz (2005) &1500,1600,1620 \\    
BX Leo & 11:38:02.1 & +16:32:36 & 11.00--11.70 &56750.782  &0.362755 &Monson  {\em et al.} (2017) &940 \\    
ST Leo & 11:38:32.7 & +10:33:42 & 10.74--12.02 & 52754.3868 &0.47799 & Maintz (2005) & 1230 \\    
Z CVn & 12:49:45.4 & +43:46:26 & 11.46--12.36  &56048.3301  & 0.6539498 &Skarka  {\em et al.} (2018), eq(1) &1690 \\    
AV Vir & 13:20:11.6 & +09:11:16 & 11.42--12.16 & 52730.3780 &0.656904 & Maintz (2005) & 1700\\    
RV UMa & 13:33:18.1 & +53:59:15 & 9.81--11.30  & 56750.455 &0.46806 & Monson  {\em et al.} (2017) &1600,1690 \\     
CN Lyr & 18:41:15.9 & +28:43:21 & 11.07--11.76 & 36079.3242 &0.41138276 &Le Borgne  {\em et al.} (2007) & 1000,1060,1065 \\    
BK Dra & 19:18:20.7 & +66:24:48 & 10.59--11.87  & 52721.6031 & 0.592076 &Maintz (2005) &1500 \\    
DM Cyg & 21:21:11.5 & +32:11:29 & 10.93--11.99  &53201.5570  &0.41986 &Maintz (2005) & 1080  \\    
AO Peg & 21:27:03.4 & +18:35:57 & 11.83--13.34  & 52876.5130 &0.54724252 & Maintz (2005) & 1090,1400,1410  \\    
BK And & 23:35:06.0 & +41:06:11 & 12.40--13.43  & 53354.3502 &0.4215985 & Maintz (2005) &1090   \\    
\enddata
\end{deluxetable}

\startlongtable 
\tabletypesize{\scriptsize}
\begin{deluxetable}{rrrrr}
\tablenum{3}
\tablewidth{0pt}
\tablecaption{Radial Velocities of RR Lyraes}
\tablehead{\colhead{Star} &\colhead{HJD-2450000} 
&\colhead{Phase} &\colhead{$V_r$} &\colhead{Sigma}\\
& & (cycles) & (km s$^{-1}$) & (km s$^{-1}$) }
\newpage
\startdata
BK And &  6909.81120 & 0.287 & -25.782 &  0.443\\
& 6909.85438 & 0.389 & $-$15.758 & 0.543\\		
& 6909.89552 & 0.487 &  $-$8.057 &  0.515\\	
& 6909.94082 & 0.594 &   1.250 &  1.251\\		
& 6910.80037 & 0.633 &   0.505 &  0.803\\		
& 6911.87746 & 0.188 & $-$37.293 &  1.035\\		
& 7408.58745 & 0.346 & $-$17.337 &  1.017\\		
& 7410.56754 & 0.043 & $-$21.932 &  1.654\\		
& 7410.60185 & 0.124 & $-$40.909 &  1.601\\		
& 7559.90291 & 0.255 & $-$32.017 &  1.288\\		
& 7562.87878 & 0.314 & $-$24.541 &  0.605\\
CI And & 6909.91591 & 0.794 & 45.224 & 1.427\\
& 6909.95995 & 0.884 & $-$1.422 & 4.122\\
& 6910.94653 & 0.920 & $-$14.619 & 1.112	\\
& 7408.61181 & 0.610 & 40.702 & 1.573\\
& 7409.60675 & 0.662 & 38.312 & 0.900	\\
& 7410.62696 & 0.767 & 46.295 & 1.189	\\
& 7410.67381 & 0.863 & 5.6990 & 3.876	\\
& 7410.69134 & 0.900 & $-$11.837 & 2.010	\\
& 7410.70868 & 0.935 & $-$19.218 & 1.332 \\
& 7410.72594 & 0.971 & $-$16.788 & 1.720\\
& 7411.66242 & 0.903 & $-$17.093 & 1.356\\
Z CVn & 7087.73866 & 0.432 & 5.596 & 1.496 \\
& 7087.82010 & 0.556 & 14.071 & 2.283 \\
& 7087.90562 & 0.687 & 17.551 & 2.064 \\
& 7091.85886 & 0.732 & 17.204 & 1.786 \\
& 7091.94005 & 0.856 & 10.689 & 1.120 \\
& 7092.71613 & 0.043 & $-$25.433 & 1.048 \\
& 7176.66366 & 0.413 &   2.280 & 1.310 \\
& 7177.65444 & 0.928 & $-$13.137 & 4.112 \\
& 7177.74755 & 0.071 & $-$15.518 & 1.910 \\
& 7408.98030 & 0.665 & 14.456 & 4.460 \\
& 7409.00708 & 0.706 & 20.755 & 6.781 \\
& 7409.90927 & 0.085 & $-$26.200 & 0.873 \\
& 7409.93213 & 0.120 & $-$20.569 & 1.231 \\
& 7409.95502 & 0.155 & $-$19.186 & 1.615 \\
& 7410.00845 & 0.237 & $-$9.861 & 0.681 \\
& 7412.91745 & 0.685 & 13.795 & 2.232 \\
& 7412.99828 & 0.809 & 18.402 & 2.402 \\
DM Cyg & 6909.63969 & 0.712 & $-$5.337 &  0.278 \\
& 6909.68283 & 0.815 & 1.870 &  0.294 \\
& 6909.72777 & 0.922 & 6.666 &  0.644 \\
& 6909.77612 & 0.037 & $-$49.487 &  1.023 \\
& 6909.83480 & 0.177 & $-$49.668 &  0.548 \\
& 6909.87379 & 0.270 & $-$37.518 &  0.472 \\
& 6910.71026 & 0.262 & $-$41.224 &  0.331 \\
& 6910.77678 & 0.420 & $-$25.609 &  0.377 \\
& 6910.83009 & 0.547 & $-$14.995 &  0.668 \\
& 6911.79925 & 0.856 & 2.922 &  0.482 \\
& 6911.85920 & 0.999 & $-$11.241 &  2.636 \\
BK Dra & 6910.63567 &  0.160 & $-$114.069 &  1.152 \\
& 6910.75071 & 0.355 & $-$93.325 &  1.194 \\
& 6911.61313 & 0.811 & $-$64.779 &  1.599 \\
& 6911.66566 & 0.900 & $-$59.485 &  1.254 \\
& 6911.82004 & 0.161 & $-$112.254 & 1.528 \\
& 7091.01239 & 0.812 & $-$66.723 &  1.392 \\
& 7092.00312 & 0.485 & $-$81.411 &  0.531 \\
& 7176.78265 & 0.675 & $-$65.473 &  1.662 \\
& 7177.81018 & 0.411 & $-$83.188 &  1.511 \\
& 7559.76185 & 0.516 & $-$76.810 &  1.652 \\
& 7559.85668 & 0.677 & $-$65.022 &  3.773 \\
& 7560.76590 & 0.212 & $-$100.154 &  1.684 \\
& 7560.86405 & 0.378 & $-$83.084 &  0.924 \\
& 7560.91625 & 0.466 & $-$74.300 &  1.040 \\
& 7561.92392 & 0.168 & $-$111.471 &  1.478 \\
& 7562.84822 & 0.729 & $-$23.021 &  1.450 \\
& 7563.76773 & 0.282 & $-$86.846 &  0.884 \\
RR Gem & 7088.64054 & 0.155 & 46.636 &  0.651 \\
& 7088.71499 & 0.343 & 64.778 &  0.593 \\
& 7088.81251 & 0.588 & 82.436 &  0.500 \\
& 7090.74639 & 0.456 & 68.287 &  0.470 \\
& 7408.75391 & 0.898 & 90.765 &  1.620 \\
& 7409.70125 & 0.282 & 56.954 &  0.211 \\
& 7409.71650 & 0.321 & 60.812 &  0.254 \\
& 7409.73241 & 0.361 & 65.182 &  0.296 \\
& 7409.82417 & 0.592 & 82.099 &  0.416 \\
& 7410.78104 & 0.000 & 22.819 &  1.919 \\
& 7410.79634 & 0.039 & 28.652 &  2.154 \\
& 7410.81476 & 0.085 & 34.977 &  0.995 \\
& 7410.82981 & 0.123 & 40.105 &  0.812 \\
& 7411.72964 & 0.388 & 66.338 &  0.263 \\
& 7411.74705 & 0.432 & 72.310 &  0.236 \\
& 7411.76195 & 0.469 & 75.276 &  0.588 \\
& 7412.73836 & 0.927 & 56.069 &  1.083 \\
& 7412.75333 & 0.965 & 33.374 &  0.939 \\
& 7412.76832 & 0.002 & 30.391 &  0.802 \\
XX Hya  & 7087.64667 & 0.362 & 70.830 &  0.946 \\
& 7087.70741 & 0.481 & 82.389 &  1.957 \\
& 7087.79318 & 0.650 & 92.007 &  0.867 \\
& 7088.69483 & 0.426 & 72.857 &  1.121 \\
& 7092.63774 & 0.191 & 49.805 &  2.493 \\
& 7408.77486 & 0.794 & 78.717 &  2.257 \\
& 7408.89992 & 0.040 & 46.429 &  4.088 \\
& 7408.91740 & 0.074 & 45.157 &  4.444 \\
& 7408.93559 & 0.110 & 50.644 &  7.781 \\
& 7409.75623 & 0.726 & 94.908 &  1.375 \\
& 7409.86736 & 0.945 & 40.797 &  5.409 \\
& 7409.88601 & 0.982 & 24.537 &  3.661 \\
& 7412.78766 & 0.696 & 93.251 &  2.352 \\
& 7412.80706 & 0.735 & 90.117 &  1.904 \\
& 7412.82631 & 0.772 & 89.560 &  2.214 \\
SS Leo  & 7177.71237 & 0.361 & 161.572 &  0.917 \\
& 7410.85218 & 0.590 & 182.033 &  1.078 \\
& 7410.87022 & 0.618 & 184.030 &  0.554 \\
& 7412.84834 & 0.777 & 184.454 &  2.348 \\
& 7413.02547 & 0.059 & 133.983 &  1.541 \\
& 7559.63506 & 0.135 & 140.159 &  1.813 \\
& 7560.65452 & 0.762 & 191.962 &  0.880 \\
& 7561.66039 & 0.368 & 164.070 &  1.064 \\
& 7562.64742 & 0.944 & 145.538 &  2.616 \\
& 7563.63426 & 0.057 & 170.456 &  0.805 \\
ST Leo & 7088.76504 & 0.927 & 131.562 &  2.899 \\
& 7090.71179 & 0.999 & 132.816 &  0.920 \\
& 7090.96678 & 0.533 & 189.935 &  1.115 \\
& 7091.81267 & 0.303 & 161.929 &  0.742 \\
& 7091.98259 & 0.658 & 195.094 &  0.853 \\
& 7092.68701 & 0.132 & 150.695 &  1.034 \\
& 7176.68903 & 0.872 & 105.257 &  1.662 \\
& 7408.95713 & 0.799 & 195.327 &  2.284 \\
& 7410.89224 & 0.847 & 145.787 &  1.758 \\
& 7410.91249 & 0.889 & 132.433 &  2.357 \\
& 7410.92971 & 0.925 & 132.761 &  2.036 \\
& 7411.97548 & 0.113 & 144.366 &  1.695 \\
& 7412.95662 & 0.166 & 157.236 &  0.753 \\
SZ Leo  &  7091.83458 & 0.612 & 172.161 &  0.817 \\
& 7091.91582 & 0.764 & 191.833 &  1.407 \\
& 7092.66430 & 0.165 & 164.103 &  2.263 \\
& 7092.76717 & 0.358 & 146.939 &  1.971 \\
& 7092.83485 & 0.485 & 160.733 &  1.080 \\
& 7176.71061 & 0.537 & 170.193 &  1.165 \\
& 7177.62628 & 0.252 & 150.505 &  13.893 \\
& 7408.80406 & 0.119 & 175.220 &  4.696 \\
& 7409.78042 & 0.948 & 186.751 &  1.677 \\
& 7409.79859 & 0.982 & 179.945 &  2.633 \\
& 7411.85812 & 0.838 & 193.081 &  4.201 \\
& 7411.87723 & 0.874 & 195.368 &  4.002 \\
BX Leo & 7091.89565 & 0.342 & 9.983 &  0.736 \\
& 7091.96354 & 0.529 & 20.096 &  1.487 \\
& 7092.74639 & 0.687 & 17.380 &  1.667 \\
& 7092.81273 & 0.870 & 12.828 &  1.952 \\
& 7092.85461 & 0.985 & $-$4.374 &  1.757 \\
& 7092.90762 & 0.131 & $-$0.608 &  0.917 \\
& 7176.73147 & 0.207 & $-$0.297 &  1.130 \\
& 7409.84477 & 0.826 & $-$1.108 &  1.207 \\
& 7410.99689 & 0.002 & 0.322 &  1.349 \\
& 7411.01496 & 0.052 & 2.010 &  0.962 \\
& 7412.87002 & 0.166 & 5.972 &  1.649 \\
& 7412.88389 & 0.204 & 7.637 &  1.797 \\
& 7412.89774 & 0.242 & 12.620 &  1.492 \\
& 7412.93947 & 0.357 & 15.305 &  1.737 \\
& 7412.97588 & 0.458 & 20.063 &  1.915 \\
& 7562.67039 & 0.118 & $-$1.252 &  0.823 \\
& 7563.66716 & 0.865 & 0.428 &  2.743 \\
TT Lyn & 6732.70866 & 0.735 & $-$46.223 & 0.763\\
CN Lyr & 6730.99650 & 0.627 & 38.301 &  0.483 \\
& 6909.66044 & 0.928 &  6.664 &  0.420 \\
& 6909.70298 & 0.031 & $-$1.092 &  0.368 \\
& 6909.75345 & 0.154 &  8.473 &  0.325 \\
& 6910.61458 & 0.247 & 14.318 &  0.341 \\
& 6910.66036 & 0.358 & 22.933 &  0.261 \\
& 6910.72864 & 0.524 & 33.791 &  0.285 \\
& 6911.69205 & 0.866 & 32.765 &  0.576 \\
& 6911.75190 & 0.012 & $-$1.294 &  0.521 \\
& 7176.96714 & 0.704 & 42.987 &  0.327 \\
& 7177.86767 & 0.893 & 18.719 &  0.807 \\
AO Peg & 6910.68824 & 0.823 & $-$247.488 &  1.407 \\
& 6911.64449 & 0.571 & $-$255.952 &  1.003 \\
& 6911.71592 & 0.701 & $-$252.748 &  1.305 \\
& 6911.77955 & 0.818 & $-$249.870 &  1.145 \\
& 7176.85660 & 0.204 & $-$292.403 &  1.652 \\
& 7176.91219 & 0.306 & $-$281.671 &  0.815 \\
& 7176.94290 & 0.362 & $-$275.285 &  0.758 \\
& 7177.88558 & 0.085 & $-$295.486 &  7.554 \\
& 7559.88541 & 0.130 & $-$299.044 &  1.585 \\
& 7560.89291 & 0.971 & $-$287.820 &  5.209 \\
& 7561.90017 & 0.811 & $-$253.902 &  1.444 \\
& 7562.82759 & 0.506 & $-$253.788 &  1.096 \\
& 7563.82565 & 0.330 & $-$277.947 &  1.089 \\
TU Per & 8389.92886 & 0.845 & $-$305.449 & 20.454\\
& 8390.87634 & 0.406 & $-$302.519 & 1.063\\
& 8390.93646 & 0.505 & $-$295.081 & 1.712\\
U Tri & 6910.97892 & 0.020 & $-$67.922 &  1.575\\
& 6911.89736 & 0.073 & $-$62.493 &  0.878\\
& 7409.62961 & 0.939 & $-$73.453 &  2.286\\
& 7409.64790 & 0.979 & $-$73.178 &  2.265\\
& 7409.66639 & 0.021 & $-$67.852 &  1.883\\
& 7410.62961 & 0.174 & $-$41.338 &  0.533\\
& 7410.75155 & 0.447 & $-$19.139 &  0.619\\
& 7411.60920 & 0.365 & $-$32.171 &  0.498\\
& 7411.63219 & 0.416 & $-$27.878 &  1.086\\
& 7411.68328 & 0.530 & $-$20.193 &  0.200\\
& 7411.70113 & 0.570 & $-$17.952 &  2.106\\
RV UMa & 7176.82601 & 0.932 & $-$191.107 &  4.308 \\
& 7176.88234 & 0.053 & $-$215.030 &  2.327 \\
& 7177.68019 & 0.757 & $-$170.711 &  0.893 \\
& 7177.83918 & 0.097 & $-$207.280 &  1.342 \\
& 7409.03257 & 0.037 & $-$208.552 &  2.036 \\
& 7410.03334 & 0.175 & $-$203.302 &  0.509 \\
& 7559.72844 & 0.995 & $-$212.169 &  3.244 \\
& 7560.72690 & 0.128 & $-$211.178 &  1.084 \\
& 7561.72147 & 0.253 & $-$200.443 &  0.567 \\
& 7562.73181 & 0.412 & $-$186.800 &  0.776 \\
& 7563.72178 & 0.527 & $-$178.679 &  0.543 \\
TU UMa & 6727.70422 & 0.960 & 72.460 &  1.118 \\
& 6727.72674 & 0.000 & 59.546 &  0.681 \\
& 6727.75031 & 0.042 & 59.482 &  1.315 \\
& 6727.77232 & 0.082 & 62.713 &  1.144 \\
& 6727.81355 & 0.156 & 70.099 &  1.040 \\
& 6727.83397 & 0.192 & 74.246 &  1.044 \\
& 6727.85827 & 0.236 & 78.348 &  0.763 \\
& 6727.88024 & 0.275 & 81.940 &  0.796 \\
& 6727.90326 & 0.317 & 85.178 &  1.091 \\
& 6727.92389 & 0.354 & 91.215 &  0.634 \\
& 6727.94718 & 0.395 & 94.446 &  0.734 \\
& 6728.64706 & 0.650 & 112.270 &  0.765 \\
& 6728.66834 & 0.689 & 114.652 &  0.452 \\
& 6728.69203 & 0.731 & 113.543 &  0.583 \\
& 6728.71282 & 0.768 & 115.355 &  0.724 \\
& 6728.73799 & 0.814 & 118.819 &  0.693 \\
& 6728.76027 & 0.853 & 121.011 &  0.854 \\
& 6729.65553 & 0.459 & 103.294 &  0.636 \\
& 6729.67593 & 0.495 & 106.113 &  0.665 \\
& 6729.70507 & 0.548 & 110.822 &  0.645 \\
& 6729.72576 & 0.585 & 113.626 &  0.684 \\
& 6729.74587 & 0.621 & 116.802 &  1.002 \\
& 6729.76732 & 0.659 & 117.818 &  0.950 \\
& 6729.80193 & 0.721 & 113.000 &  0.836 \\
& 6729.82691 & 0.766 & 113.816 &  0.510 \\
& 6729.84919 & 0.806 & 116.342 &  0.486 \\
& 6729.87322 & 0.849 & 119.478 &  0.641 \\
& 6729.89742 & 0.893 & 119.896 &  0.997 \\
& 6729.92019 & 0.933 & 93.539 &  2.912 \\
& 6729.95564 & 0.997 & 61.490 &  1.141 \\
& 6729.97776 & 0.037 & 59.359 &  0.560 \\
& 6729.99928 & 0.075 & 62.363 &  1.396 \\
& 6730.64723 & 0.237 & 77.273 &  0.740 \\
& 6730.66868 & 0.276 & 81.912 &  0.569 \\
& 6730.69088 & 0.315 & 86.302 &  0.586 \\
& 6730.71392 & 0.357 & 90.731 &  0.388 \\
& 6730.73419 & 0.393 & 94.003 &  0.350 \\
& 6730.75439 & 0.429 & 97.074 &  0.385 \\
& 6730.77483 & 0.466 & 100.348 &  0.420 \\
& 6730.79585 & 0.504 & 103.863 &  0.409 \\
& 6730.83230 & 0.569 & 109.607 &  0.458 \\
& 6730.85561 & 0.611 & 112.521 &  0.682 \\
& 6730.88421 & 0.662 & 114.452 &  0.856 \\
& 6730.90685 & 0.703 & 113.878 &  0.615 \\
& 6730.92668 & 0.738 & 114.790 &  0.515 \\
& 6730.95985 & 0.798 & 114.359 &  0.931 \\
& 6730.97978 & 0.834 & 117.274 &  0.625 \\
& 7087.68335 & 0.478 & 103.435 &  0.678 \\
& 7087.76975 & 0.633 & 114.920 &  0.729 \\
& 7087.85302 & 0.783 & 116.884 &  0.916 \\
& 7087.93603 & 0.932 & 93.129 &  2.460 \\
& 7087.97139 & 0.995 & 63.872 &  0.954 \\
& 7088.66909 & 0.246 & 81.292 &  0.399 \\
& 7088.74142 & 0.376 & 94.091 &  0.299 \\
& 7088.78689 & 0.457 & 100.469 &  0.469 \\
AV Vir  & 7038.96697 & 0.932 & 172.306 & 2.049\\
& 7038.98836 & 0.965 & 145.256 & 1.814\\
& 7039.00899 & 0.996 & 132.296 & 1.759\\
& 7087.87912 & 0.391 & 150.374 &  0.531 \\
& 7087.99680 & 0.570 & 162.661 &  0.934 \\
& 7090.98956 & 0.126 & 120.879 &  1.053 \\
& 7092.79103 & 0.868 & 173.648 &  0.997 \\
& 7092.87556 & 0.997 & 130.487 &  1.207 \\
& 7092.92858 & 0.078 & 123.623 &  0.809 \\
& 7176.75669 & 0.689 & 164.730 &  0.711 \\
& 7177.77775 & 0.243 & 142.178 &  1.061 \\ 
& 7409.98005 & 0.723 & 169.815 &  1.119 \\
& 7410.95077 & 0.201 & 132.762 &  0.723 \\
& 7411.03626 & 0.331 & 145.101 &  0.708 \\
\enddata
\end{deluxetable} 
                                          
\begin{deluxetable}{rcrcl}
\tablenum{4}
\tablewidth{0pt}
\tablecaption{New Center-of-Mass Radial Velocities}
\tablehead{\colhead{Star} & \colhead{JD-2450000} & \colhead{Velocity} & \colhead{Uncertainty} & \colhead{Type of Fit}\\
& & (km s$^{-1}$) & (km s$^{-1}$) }
\newpage
\startdata
BK And & 7164.9743 & $-$5.98 & 1.07 & Template\\ 
CI And & 7274.0118 & 19.15 & 0.94 & Template\\
Z CVn & 7256.2207 & $-$0.55 & 0.67 & Fourier Series \\ 
DM Cyg & 6910.4100 & $-$21.05 & 0.84 & Template \\ 
BK Dra & 7313.3672 & $-$85.34 & 1.12 & Template \\ 
RR Gem & 7343.2296 & 65.48 & 0.74 & Template \\
XX Hya & 7303.1958& 65.81 & 1.11 & Fourier Series \\
SS Leo & 7463.3540 & 163.63 & 0.49 & Wtd. Mean \\ 
ST Leo & 7245.4029 & 164.90 & 2.54 & Template \\ 
SZ Leo & 7239.0393 & 171.31 & 0.75 & Fourier Series \\
BX Leo & 7303.2160 & 9.24 & 0.79 & Wtd. Mean\\ 
TT Lyn & 6732.7086 & $-$63.25 & \ldots  & Template \\
CN Lyr & 6942.7632 & 23.62 & 0.21 & Wtd. Mean \\ 
AO Peg & 7243.3659 & $-$271.79 & 0.68 & Fourier Series \\ 
TU Per & 8390.5805 & $-$308.56 & 1.84 & Template \\
U Tri & 7319.9842 & $-$40.05 & 1.27 & Template \\
RV UMa & 7394.2658 & $-$187.84 & 0.74 & Wtd. Mean \\
TU UMa & 6781.6953 & 94.15 & 0.17 & Fourier Series \\ 
AV Vir & 7160.6376 & 151.47 & 0.66 & Wtd. Mean\\ 
\enddata
\end{deluxetable}

\startlongtable 
\tabletypesize{\scriptsize}
\begin{deluxetable}{crrrrl}
\tablenum{5}
\tablewidth{0pt}
\tablecaption{Historical Center--of--Mass Radial Velocities}
\tablehead{\colhead{Star} &\colhead{Julian Date} 
&\colhead{Gamma} &\colhead{Sigma} &\colhead{Residual} &\colhead{Source}\\
& & (km s$^{-1}$) & (km s$^{-1}$) & (km s$^{-1}$) }
\newpage
\startdata
BK And & 2448133 & 117.53 &  47.00 & 127.00 & Layden (1994)\\
& 2450335 & $-$9.68 & 0.26 & -0.21 & Jeffery {\em et al.} (2007)\\
& 2457165 & $-$5.98 & 1.07 & 3.49 & This work\\
& Wtd. Mean & $-$9.47&0.25 & &\\
CI And & 2448131 & 99.00 & 30.00 & 79.26 & Layden (1994)\\
& 2449488 & 23.00 & 3.00 & 3.26 & Fernley \& Barnes (1997)\\
& 2453185 & 20.07  & 5.00 & 0.33 & S. Liu {\em et al.} (2013)\\
& {\em 2455197} & {\em 51.21} & {\em 12.14} & {\em 31.47} & G.--C. Liu {\em et al.} (2020)\\
& 2457274 & 19.15 & 0.94 & $-$0.59 & This work\\
& Wtd. Mean & 19.74 & 0.88 & &\\
Z CVn & 2423535 & 8.10 & 13.00 & 8.40 & Joy (1938); Skarka {\em et al.} (2018)\\
& 2445462 & 10.60 & 4.10 & 10.90 & Hawley \& Barnes (1985), Skarka {\em et al.} (2018)\\
& 2447974 & $-$3.00 & 26.00 & $-$2.70 & Layden (1994)\\
& 2449473 & 11.00 & 10.00 & 11.30 & Fernley \& Barnes (1997)\\
& {\em 2455197} &  {\em $-$13.49} &  {\em 7.24} &  {\em $-$13.20} & G.--C. Liu {\em et al.} (2020)\\
& 2457256 & $-$0.55 & 0.67 & $-$0.25 & This work\\
& Wtd. Mean &	$-$0.30 & 0.66 & &\\
DM Cyg & 2429121 & $-$49.00 & 28.00 & $-$27.21 & Joy (1938), Payne--Gaposhkin (1954)\\
& 2448131 & 59.00 & 30.00 & 80.79 & Layden (1994)\\
& 2449519 & $-$35.00 & 3.00 & $-$13.21 & Fernley \& Barnes (1997)\\
& 2450301 & $-$20.52 & 1.90 & 1.27 & Jeffery {\em et al.} (2007)\\
& 2453185 & $-$21.60 & 5.00 & 0.19 & S. Liu {\em et al.} (2013)\\
& 2456910 & $-$21.05 & 0.84 & 0.74 & This work\\
& Wtd. Mean & $-$21.79 & 0.74 & &  \\
BK Dra & 2447289 & $-$28.00 & 30.00 & 57.22 & Layden (1994)\\
& 2449473 & $-$82.00 & 10.00 & 3.22 & Fernley \& Barnes (1997)\\
& 2457313 & $-$85.34 & 1.12 & $-$0.12 & This work\\
& Wtd. Mean & $-$85.22 & 1.11 &  &\\
RR Gem & 2424532 & 94.00 & 28.00 & 28.43 & Joy (1938), Payne--Gaposhkin (1954)\\
& 2445302 & 60.00 & 5.00 & $-$5.57 & Hawley \& Barnes (1985)\\
& 2447220 & 63.90 & 7.34 & $-$1.67 & Liu \& Janes (1989)\\
& 2447973 & 46.00 & 26.00 & -19.57 & Layden (1994)\\
& {\em 2455197}  &  {\em 78.37} &  {\em 5.53} &  {\em 12.79} & G.--C. Liu {\em et al.} (2020)\\
& 2457343 & 65.48 & 0.74 & $-$0.09 & This work\\
& Wtd. Mean & 65.57 & 0.72 & &\\
XX Hya &  {\em 2431091} &  {\em $-$10.00} &  {\em 28.00} &  {\em $-$75.97} & Joy (1950)\\
& 2447473 & 52.00 & 24.00 & $-$13.97 & Layden (1994)\\
& 2449488 & 95.00 & 10.00 & 29.03 & Ferney \& Barnes (1997)\\
& 2449701 & 32.00 & 30.00 & $-$33.97 & Solano {\em et al.} (1997)\\
& 2457303 & 65.81 & 1.11 & $-$0.16 & This work\\
& Wtd. Mean & 65.97 & 1.10 &  & \\
SS Leo & 2427845 & 145.00 & 28.00 & $-$16.74 & Joy (1938), Payne--Gaposhkin (1954)\\
& 2445425 & 180.00 & 7.50 & 18.26 & Hawley \& Barnes (1985)\\
& 2445841 & 160.70 & 0.35 & $-$1.04 & Fernley {\em et al.} (1990)\\
& 2446511 & 162.50 & 6.80 & 0.76 & Carrillo {\em et al.} (1995)\\
& 2447524 & 161.00 & 18.00 & $-$0.74 & Layden (1994)\\
& 2454951 & 160.63 & 3.26 & $-$1.12 & Britavskiy {\em et al.} (2018)\\
& {\em 2455197}  &  {\em 177.42} &  {\em 5.53} &  {\em 15.68} & G.--C. Liu {\em et al.} (2020)\\
& 2455578 & 162.00 & 1.50 & 0.26 & Chadid {\em et al.} (2017)\\
& 2457463 & 163.63 & 0.50 & 1.88 & This work\\
& Wtd. Mean & 161.74 & 0.28 & &\\
ST Leo & 2427454 & 170.00 & 35.00 & 3.92 & Joy (1938), Payne--Gaposhkin (1954)\\
& {\em 2436205} &  {\em 150.00} &  {\em 4.00} &  {\em $-$16.08} & Rogers (1960)\\
& {\em 2438799} &  {\em 166.00} &  {\em 10.00} &  {\em $-$0.08} & Woolley \& Savage (1971)\\
& 2447235 & 181.00 & 39.00 & 14.92 & Layden (1994)\\
& 2449473 & 177.00 & 3.00 & 10.92 & Fernley \& Barnes (1997)\\
& 2455656 & 166.00 & 1.50 & $-$0.08 & Chadid {\em et al.}  (2017)\\
& 2457245 & 164.90 & 2.54 & $-$1.18 & This work\\
& Wtd. Mean & 166.08 & 1.13 & &\\
SZ Leo & 2427078 & 90.00 & 28.00 & $-$81.01 & Joy (1938)\\
& 2445446 & 187.00 & 5.00 & 15.99 & Hawley \& Barnes (1985)\\
& 2447644 & 171.00 & 14.00 & $-$0.01 & Layden (1994)\\
& {\em 2455197}  &  {\em 147.80} &  {\em 4.67} &  {\em $-$23.21} & G.--C. Liu {\em et al.}  (2020)\\
& 2457239 & 171.32 & 0.75 & 0.30 & This work\\
& Wtd. Mean & 171.01 & 0.73 & &\\
BX Leo & 2449473 & 27.00 & 3.00 & 16.65 & Fernley \& Barnes (1997)\\
& 2449487 & $-$7.00 & 15.00 & $-$17.35 & Solano {\em et al.} (1997)\\
& 2457303 & 9.24 & 0.79 & $-$1.11 & This work\\
& Wtd. Mean & 10.35 & 0.77 & &\\
TT Lyn & 2438727 & $-$63.70 & 6.80 & $-$0.83 & Woolley \& Aly (1966)\\
& 2444473 & $-$62.70 & 2.00 & 0.17 & Barnes {\em et al.} (1988)\\
& 2446953 & $-$66.60 & 7.34 & $-$3.73 & T. Liu \& Janes (1989)\\
& 2447974 & $-$50.61 & 3.31 & 12.26 & Layden (1994)\\
& 2449488 & $-$65.00 & 3.00 & $-$2.13 & Fernley \& Barnes (1997)\\
& 2449701 & $-$50.00 & 15.00 & 12.87 & Solano {\em et al.} (1997)\\
& {\em 2455197}  &  {\em $-$86.60} &  {\em 5.53} &  {\em $-$23.73} & G.--C. Liu {\em et al.} (2020)\\
& 2456733 & $-$63.25 & 1.80 & $-$0.38 & This work\\
& Wtd. Mean & $-$62.87 & 1.09 &  &\\
CN Lyr & 2448130 & 67.00 & 30.00 & 43.36 & Layden (1994)\\
& 2449473 & 27.00 & 3.00 & 3.36 & Fernley \& Barnes (1997)\\
& 2449701 & 13.00 & 20.00 & $-$10.64 & Solano {\em et al.} (1997)\\
& 2453185 & 26.04 & 5.00 & 2.40 & S. Liu {\em et al.} (2013)\\
& 2456943 & 23.62 & 0.21 & $-$0.02 & This work\\
& Wtd. Mean & 23.64 & 0.21 & &\\
AO Peg &  2432917 & 115.00 & 28.00 &  384.25 & Joy (1950)\\
&  2448082 & $-$233.77 & 13.00 &  35.48 & Layden (1994)\\
&  2450673 & $-$268.78 & 0.11 &  0.47 & Jeffery {\em et al.} (2007)\\
&  2453185 & $-$277.56& 0.50 &  $-$8.31 & S. Liu {\em et al.}  (2013)\\
&  {\em 2455197} &  {\em $-$336.34} &  {\em 7.36} &   {\em $-$67.08} & G.--C. Liu {\em et al.}  (2013)\\
&  2457243 & $-$271.79 & 0.68 &  $-$2.54 & This work\\
&  Wtd. Mean & $-$269.25 & 0.11 &  & \\
TU Per & 2425199 & $-$380.00 & 28.00 & $-$66.14 & Joy (1938)\\
& 2445361 & $-$327.36 & 10.00 & $-$13.50 & Hawley \& Barnes (1985)\\
& 2451673 & $-$314.24 & 0.54 & $-$0.38 & Jeffery {\em et al.} (2007)\\
& {\em 2455197} &  {\em $-$315.43} &  {\em 7.36} &  {\em $-1$.58} & G.--C. Liu {\em et al.} (2020)\\
& 2458391 & $-$308.56 & 1.84 & 5.30 & This work\\
& Wtd. Mean & $-$313.86 & 0.52 &  &\\
U Tri & 2429493 & $-$60.00 & 28.00 & $-$22.14 & Joy (1955)\\
& 2448132 & 56.00 & 30.00 & 93.86 & Layden (1994)\\
& 2450354 & $-$36.73 & 0.94 & 1.13 & Jeffery {\em et al.} (2007)\\
& 2457320 & $-$40.05 & 1.27 & $-$2.19 & This work\\
& Wtd. Mean & $-$37.86 & 0.76 &  &\\
RV UMa & 2422765 & $-$178.00 & 28.00 & 9.67 & Joy (1938), Payne--Gaposhkin (1954)\\
& 2436841 & $-$181.00 & 5.00 & 6.67 & Preston \& Spinrad (1967)\\
& 2447975 & $-$195.00 & 21.00 & $-$7.33 & Layden (1994)\\
& 2448011 & $-$187.67 & 0.54 & 0.00 & Fernley {\em et al.} (1993)\\
& 2454921 & $-$186.73 & 5.01 & 0.94 & Britavskiy {\em et al.} (2018)\\
& 2457394 & $-$187.84 & 0.74 & $-$0.17 & This work\\
& Wtd. Mean & $-$187.67 & 0.43 &  &\\
TU UMa & 2426076 & 104.00 & 28.00 & 10.25 & Joy (1938), Payne--Gaposhkin (1954)\\
& 2436821 & 92.00 & 1.00 & $-$1.75 & Preston {\em et al.} (1961)\\
& 2438039 & 88.01 & 0.76 & $-$5.74 & Preston \& Paczynski (1964)\\
& 2443993 & 90.00 & 2.00 & $-$3.75 & Barnes {\em et al.}  (1988)\\
& 2446894 & 77.00 & 2.00 & $-$16.75 & Saha \& White (1990)\\
& 2446985 & 84.20 & 7.34 & $-$9.55 & T. Liu \& Janes (1989)\\
& 2447975 & 75.00 & 17.00 & $-$18.75 & Layden (1994)\\
& 2449473 & 101.00 & 3.00 & 7.25 & Fernley \& Barnes (1997)\\
& 2449487 & 96.00 & 3.00 & 2.25 & Solano {\em et al.} (1997)\\
& 2454922 & 102.95 & 2.40 & 9.20 & Britavskiy {\em et al.} (2018)\\
& 2456782 & 94.15 & 0.17 & 0.40 & This work\\
& Wtd. Mean & 93.75 & 0.16 & &\\
AV Vir & 2432689 & 35.00 & 28.00 & $-$116.43 & Joy (1950), Payne--Gaposhkin (1954)\\ 
& 2445410 & 154.00 & 5.00 & 2.57 & Hawley \& Barnes (1985)\\
& 2447246 & 119.00 & 29.00 & $-$32.43 & Layden (1994)\\
& 2457161 & 151.47 & 0.66 & 0.04 & This work\\
& Wtd. Mean & 151.43 & 0.65 & &\\
\enddata
\end{deluxetable}

\begin{figure}
\figurenum{1}
  \epsscale{0.95}
    \includegraphics[angle=0, width=6.5in]{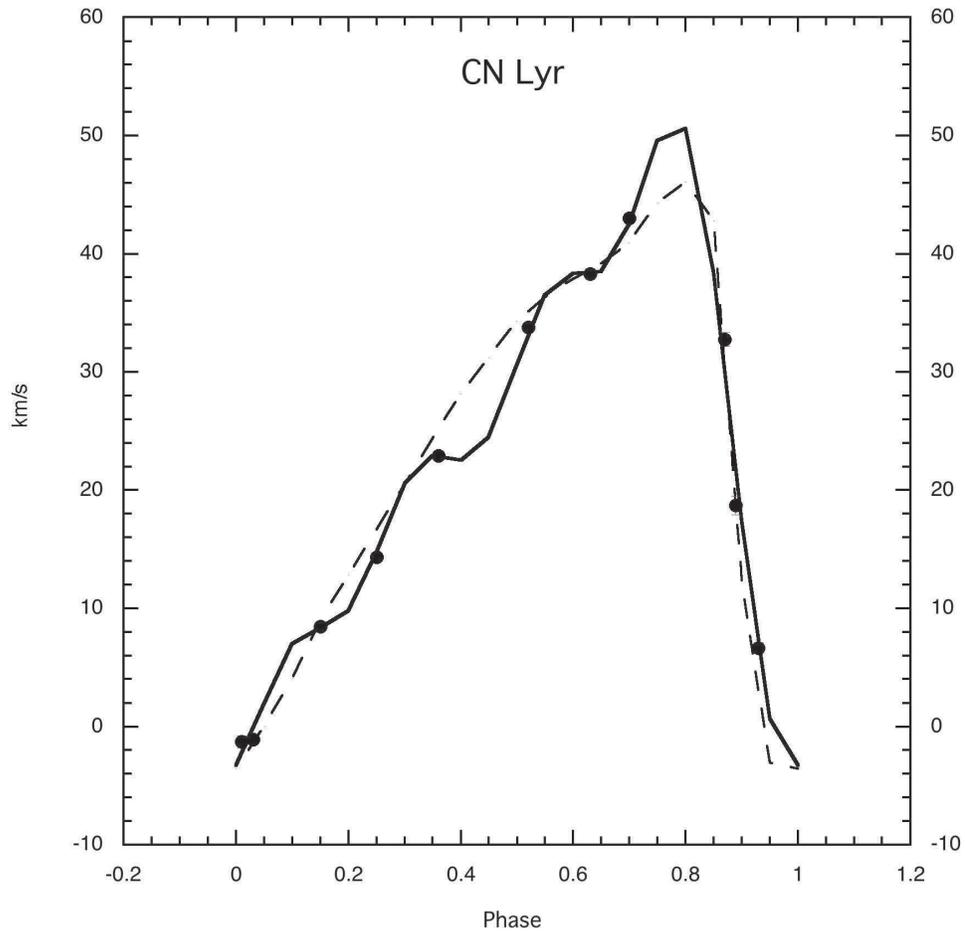}
      \figcaption{A comparison of the Fourier Series (order = 4) and Template (phase shift = $-$0.04) fits to the CN Lyr data. The FS is the solid curve and the Template is the dashed curve. The observed radial velocities are shown as filled circles, with error bars typically smaller than the symbol. Well within the uncertainties, both fits give the same gamma velocity.}
   \label{Fig1}
\end{figure}

\begin{figure}
\figurenum{2}
  \epsscale{0.95}
    \includegraphics[angle=0, width=6.5in]{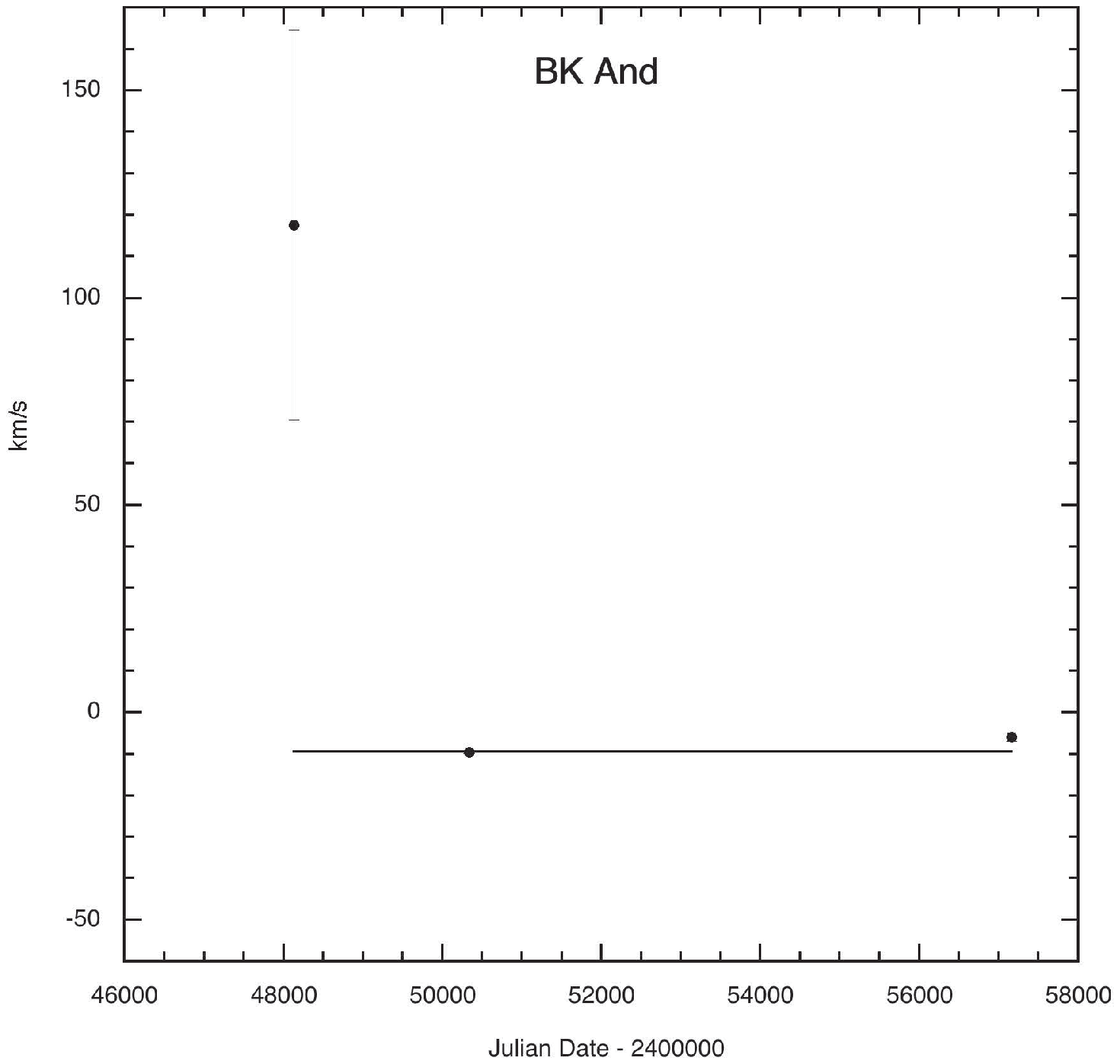}
      \figcaption{Center--of--mass radial velocities for BK And. Some error bars are smaller than the points plotted. The horizontal line shows the weighted mean radial velocity.}
   \label{Fig2}
\end{figure}

\begin{figure}
\figurenum{3}
  \epsscale{0.95}
    \includegraphics[angle=0, width=6.5in]{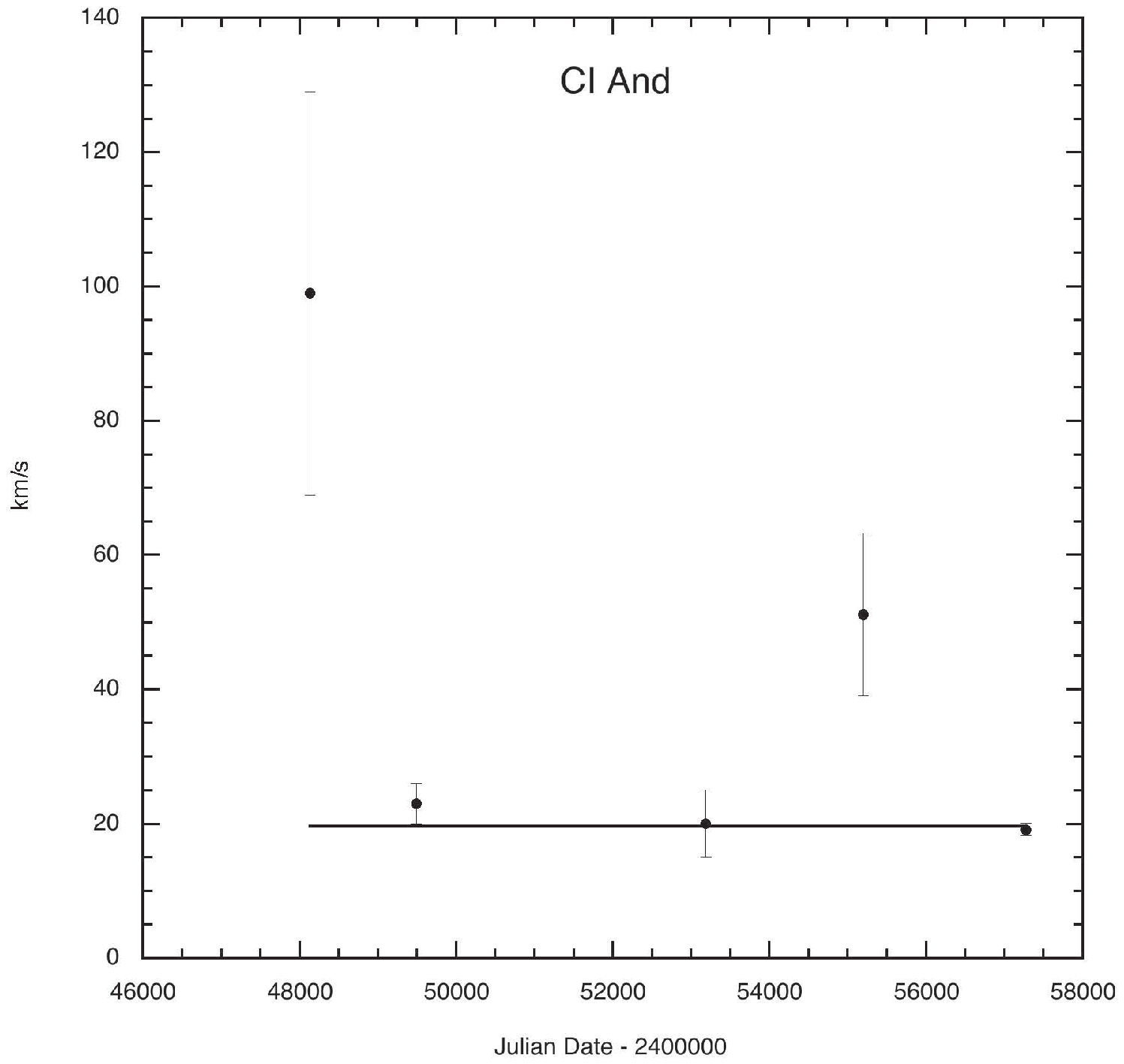}
      \figcaption{Center--of--mass radial velocities for CI And. The datum at 55197 for G.--C. Liu {\em et al.} is illustrative only for the JD (see text). Some error bars are smaller than the points plotted. The horizontal line shows the weighted mean radial velocity.}
   \label{Fig3}
\end{figure}

\begin{figure}
\figurenum{4}
  \epsscale{0.95}
    \includegraphics[angle=0, width=6.5in]{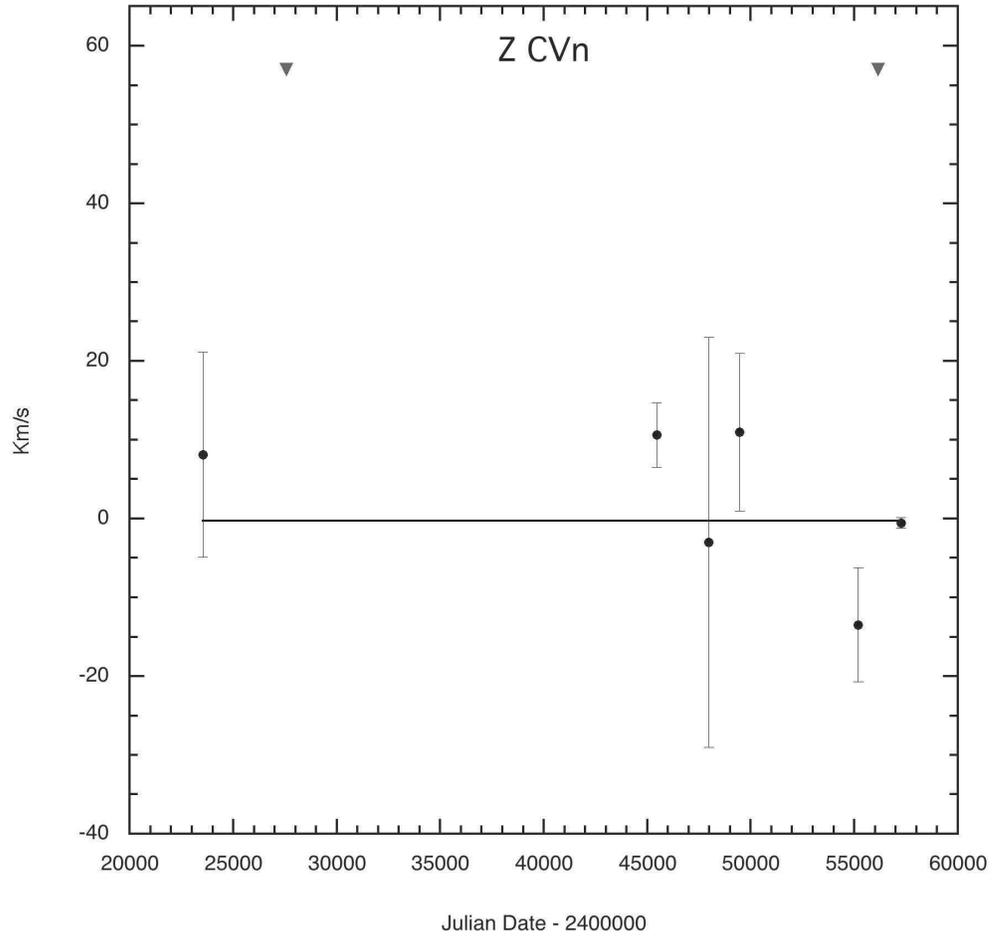}
      \figcaption{Center--of--mass radial velocities for Z CVn. The datum at 55197 for G.--C. Liu {\em et al.} is illustrative only for the JD (see text). The horizontal line shows the weighted mean radial velocity.  The predicted dates of periastron passage and the periastron center--of--mass reflex radial velocity for the 78.3 year orbit are shown as inverted filled triangles.}
   \label{Fig4}
\end{figure}

\begin{figure}
\figurenum{5}
  \epsscale{0.95}
    \includegraphics[angle=0, width=6.5in]{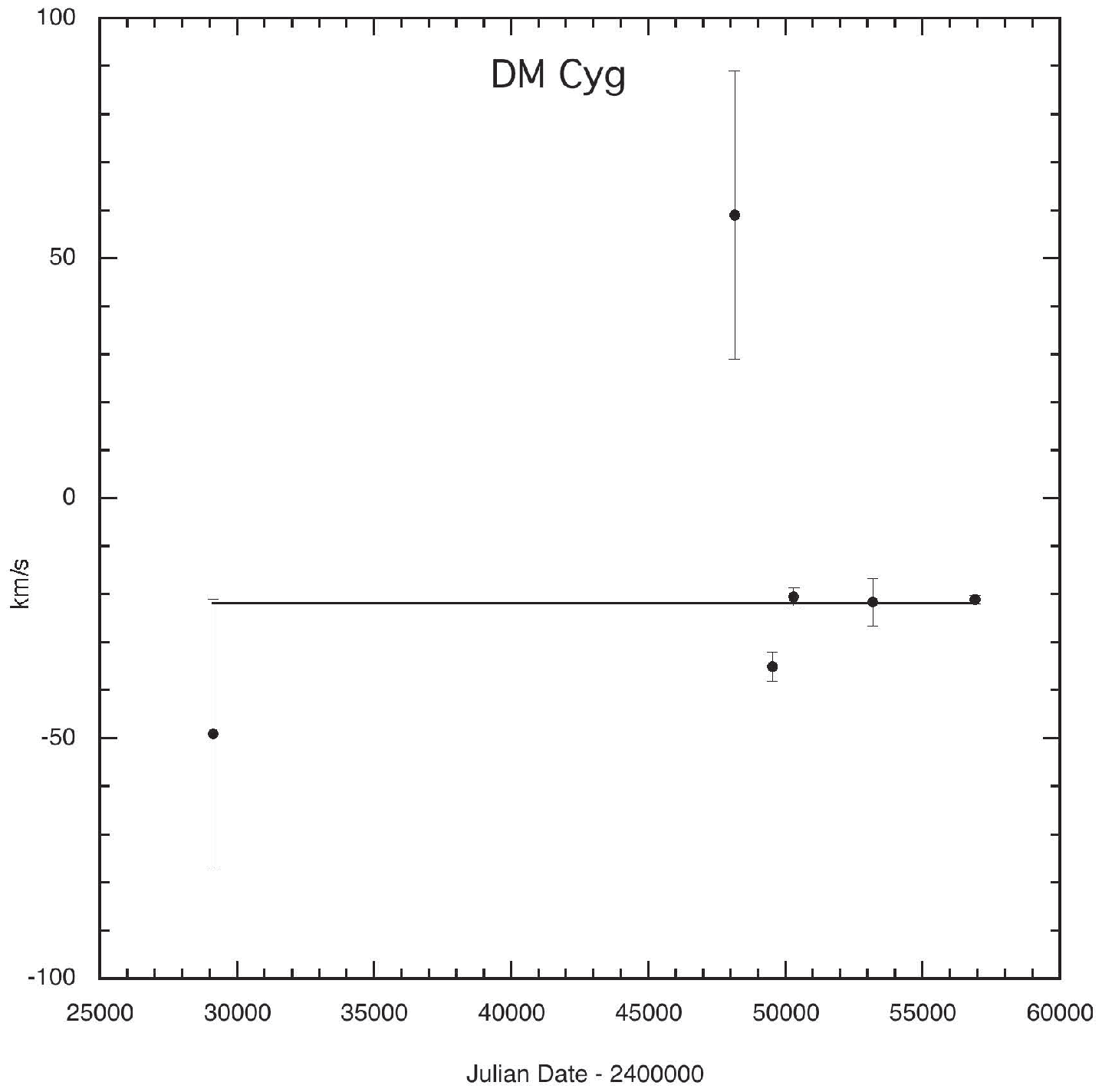}
      \figcaption{Center--of--mass radial velocities for DM Cyg. Some error bars are smaller than the points plotted. The horizontal line shows the weighted mean radial velocity.}
   \label{Fig5}
\end{figure}

\begin{figure}
\figurenum{6}
  \epsscale{0.95}
    \includegraphics[angle=0, width=6.5in]{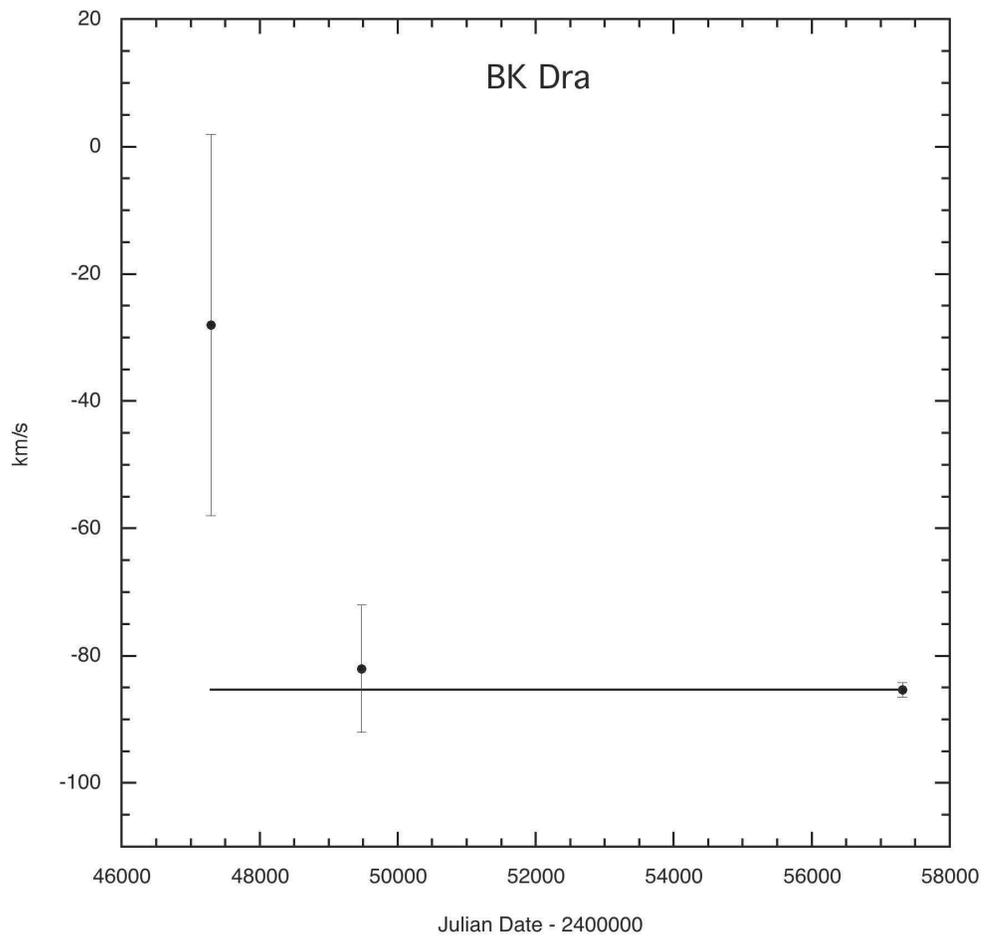}
      \figcaption{Center--of--mass radial velocities for BK Dra. The horizontal line shows the weighted mean radial velocity.}
   \label{Fig6}
\end{figure}

\begin{figure}
\figurenum{7}
  \epsscale{0.95}
    \includegraphics[angle=0, width=6.5in]{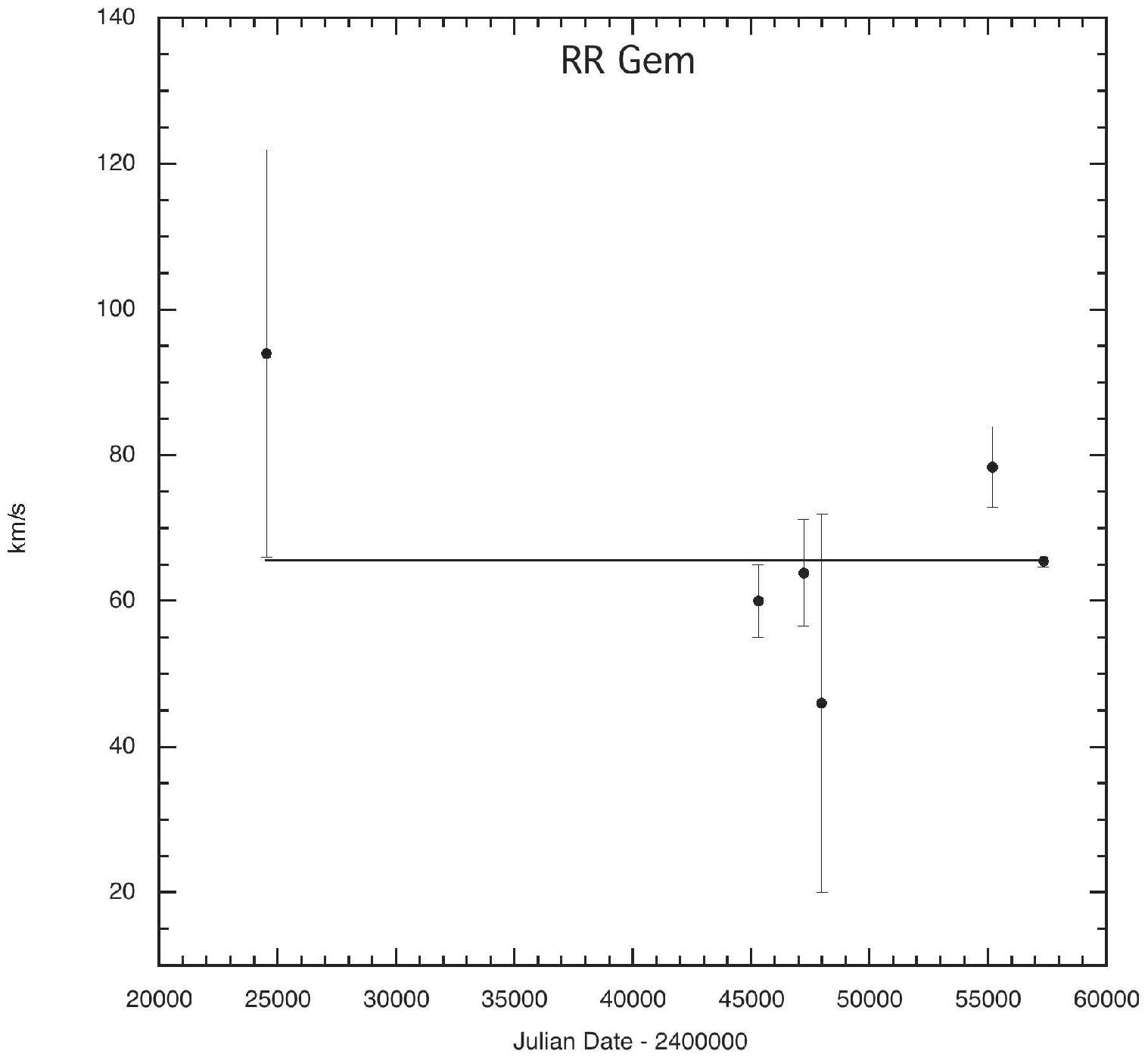}
      \figcaption{Center--of--mass radial velocities for RR Gem. The datum at 55197 for G.--C. Liu {\em et al.} is illustrative only for the JD (see text). Some error bars are smaller than the points plotted. The horizontal line shows the weighted mean radial velocity.}
   \label{Fig7}
\end{figure}

\begin{figure}
\figurenum{8}
  \epsscale{0.95}
    \includegraphics[angle=0, width=6.5in]{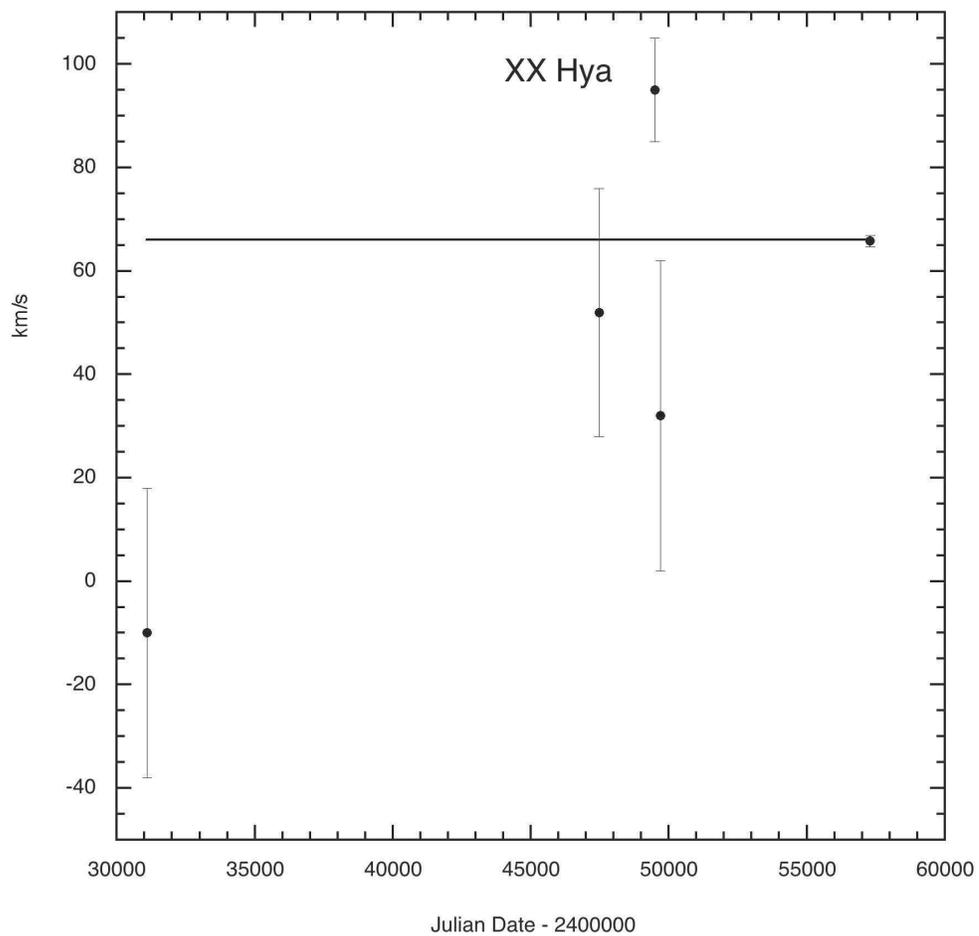}
      \figcaption{Center--of--mass radial velocities for XX Hya. The datum at 31091 for Joy (1950) is illustrative only for the JD (see text). Some error bars are smaller than the points plotted. The horizontal line shows the weighted mean radial velocity.}
   \label{Fig8}
\end{figure}

\begin{figure}
\figurenum{9}
  \epsscale{0.95}
    \includegraphics[angle=0, width=6.5in]{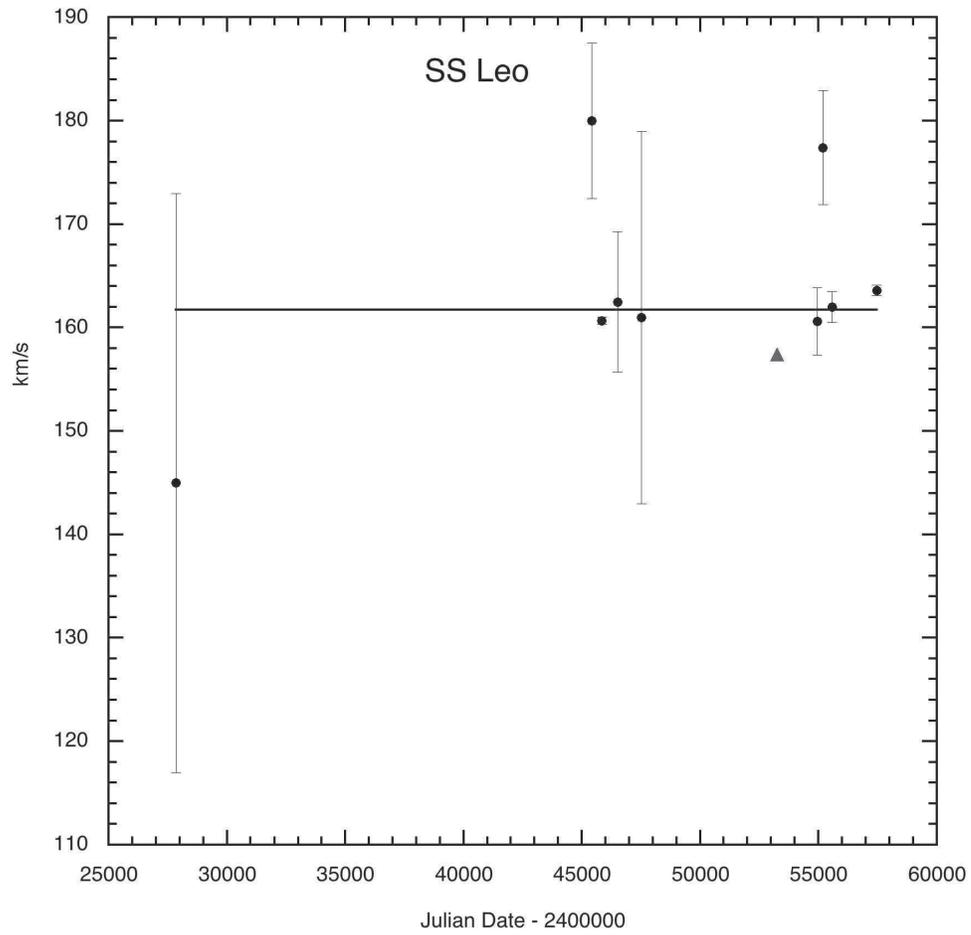}
      \figcaption{Center--of--mass radial velocities for SS Leo. The datum at 55197 for G.--C. Liu {\em et al.} is illustrative only for the JD (see text). Some error bars are smaller than the points plotted. The horizontal line shows the weighted mean radial velocity. The predicted date of periastron passage and the periastron center--of--mass reflex radial velocity for the 110.7 year orbit is shown as a filled triangle.}
   \label{Fig9}
\end{figure}

\begin{figure}
\figurenum{10}
  \epsscale{0.95}
    \includegraphics[angle=0, width=6.5in]{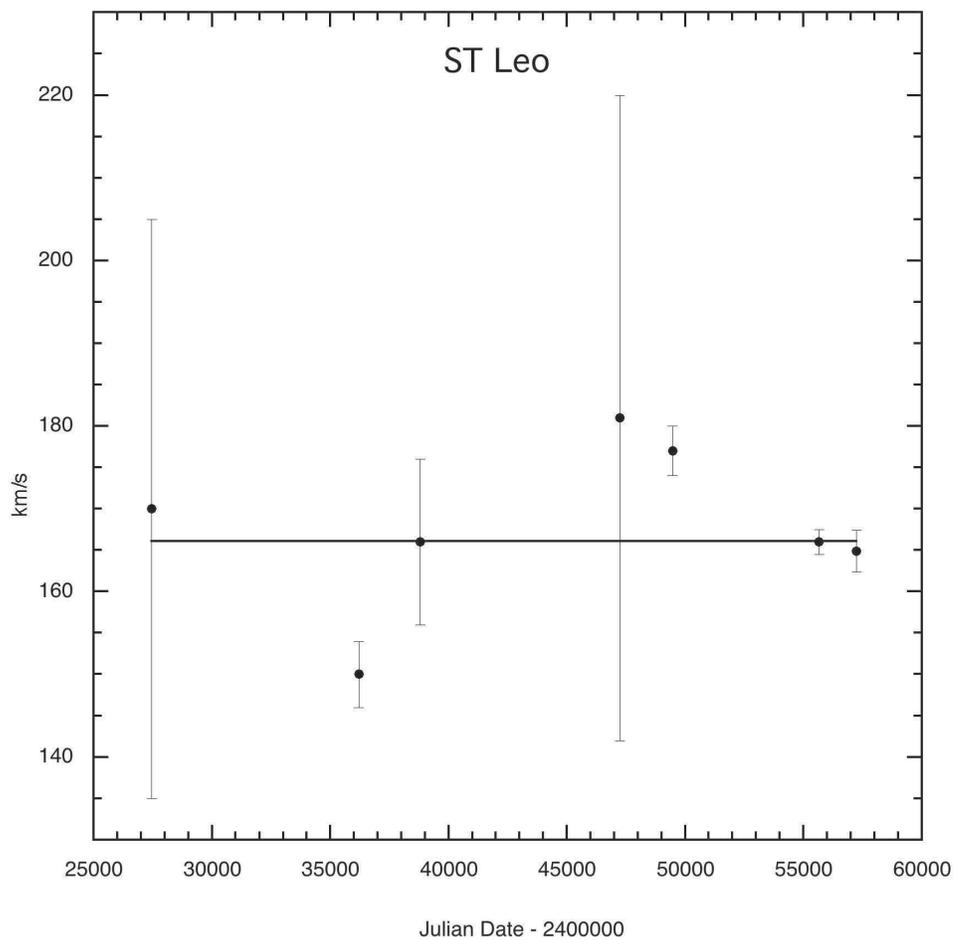}
      \figcaption{Center--of--mass radial velocities for ST Leo. The data at 36205 for Rogers (1960) and at 38799 for Woolley \& Savage (1971) are illustrative only for the JD (see text). The horizontal line shows the weighted mean radial velocity.}
   \label{Fig10}
\end{figure}

\begin{figure}
\figurenum{11}
  \epsscale{0.95}
    \includegraphics[angle=0, width=6.5in]{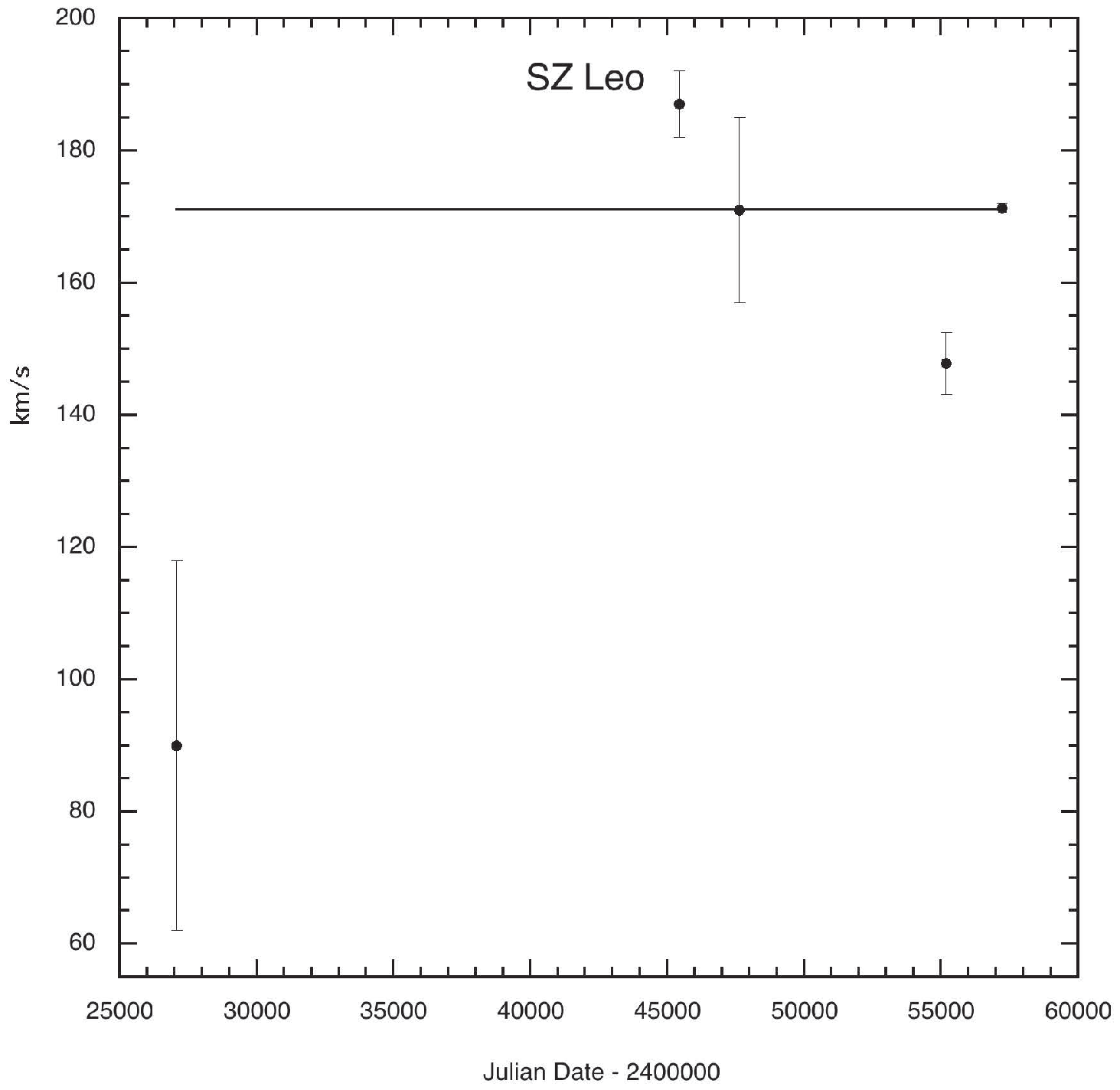}
      \figcaption{Center--of--mass radial velocities for SZ Leo. The datum at 55197 for G.--C. Liu {\em et al.} is illustrative only for the JD (see text). Some error bars are smaller than the points plotted. The horizontal line shows the weighted mean radial velocity.}
   \label{Fig11}
\end{figure}

\begin{figure}
\figurenum{12}
  \epsscale{0.95}
    \includegraphics[angle=0, width=6.5in]{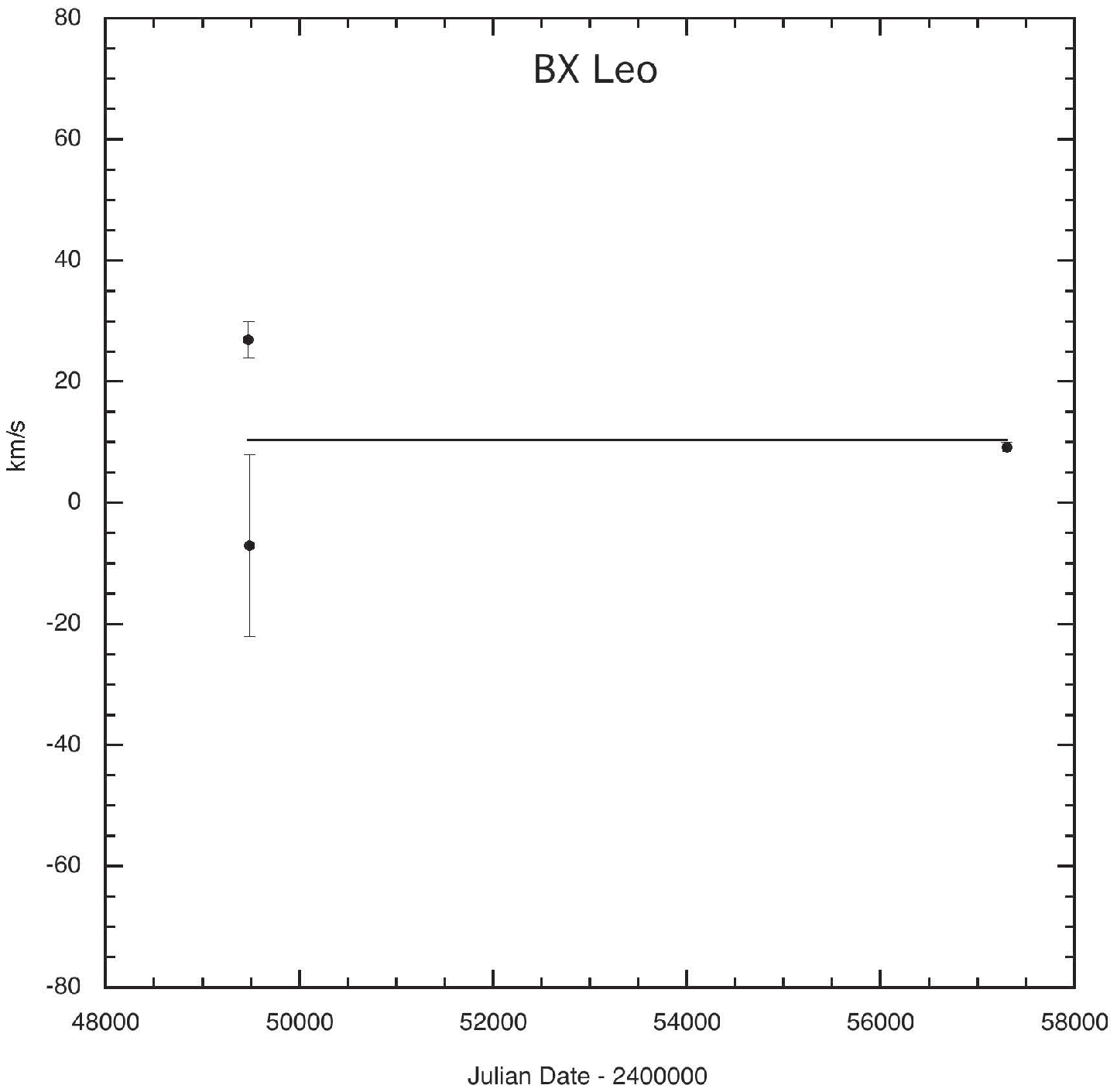}
      \figcaption{Center--of--mass radial velocities for BX Leo. Some error bars are smaller than the points plotted. The horizontal line shows the weighted mean radial velocity.}
   \label{Fig12}
\end{figure}

\begin{figure}
\figurenum{13}
  \epsscale{0.95}
    \includegraphics[angle=0, width=6.5in]{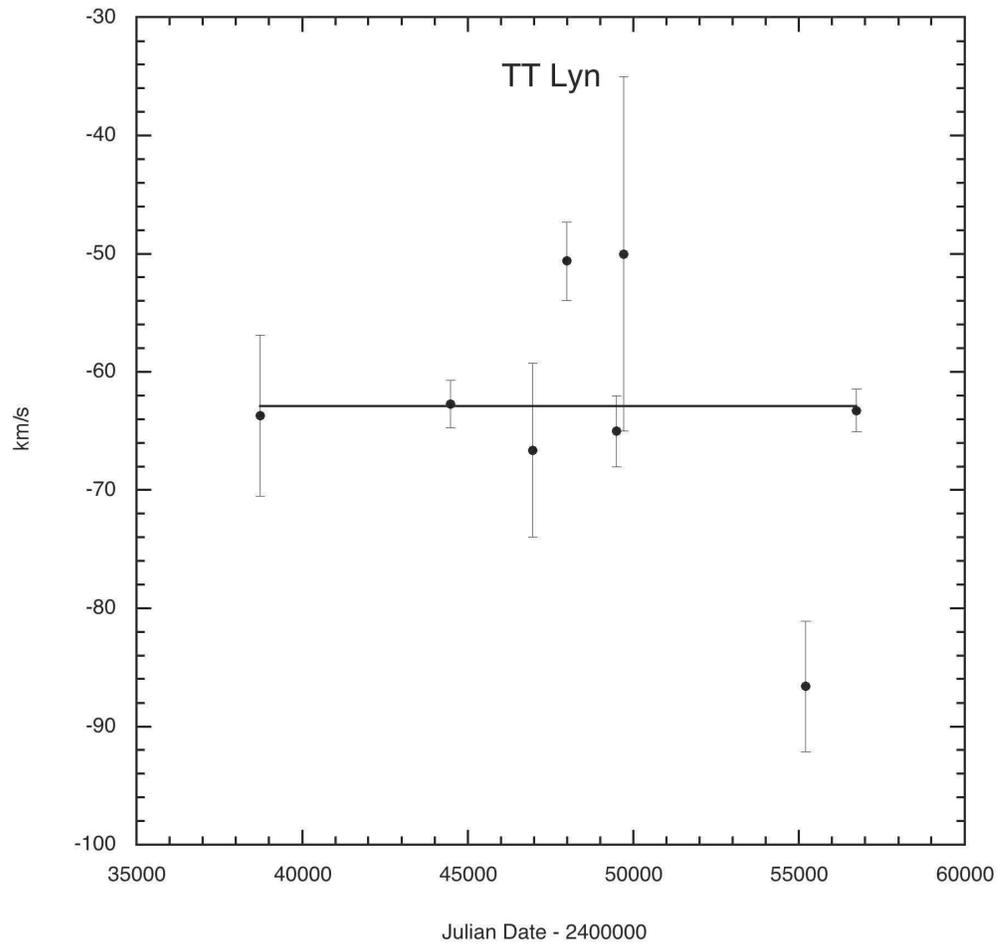}
      \figcaption{Center--of--mass radial velocities for TT Lyn. The datum at 55197 for G.--C. Liu {\em et al.} is illustrative only for the JD (see text).  The horizontal line shows the weighted mean radial velocity.}
   \label{Fig13}
\end{figure}

\begin{figure}
\figurenum{14}
  \epsscale{0.95}
    \includegraphics[angle=0, width=6.5in]{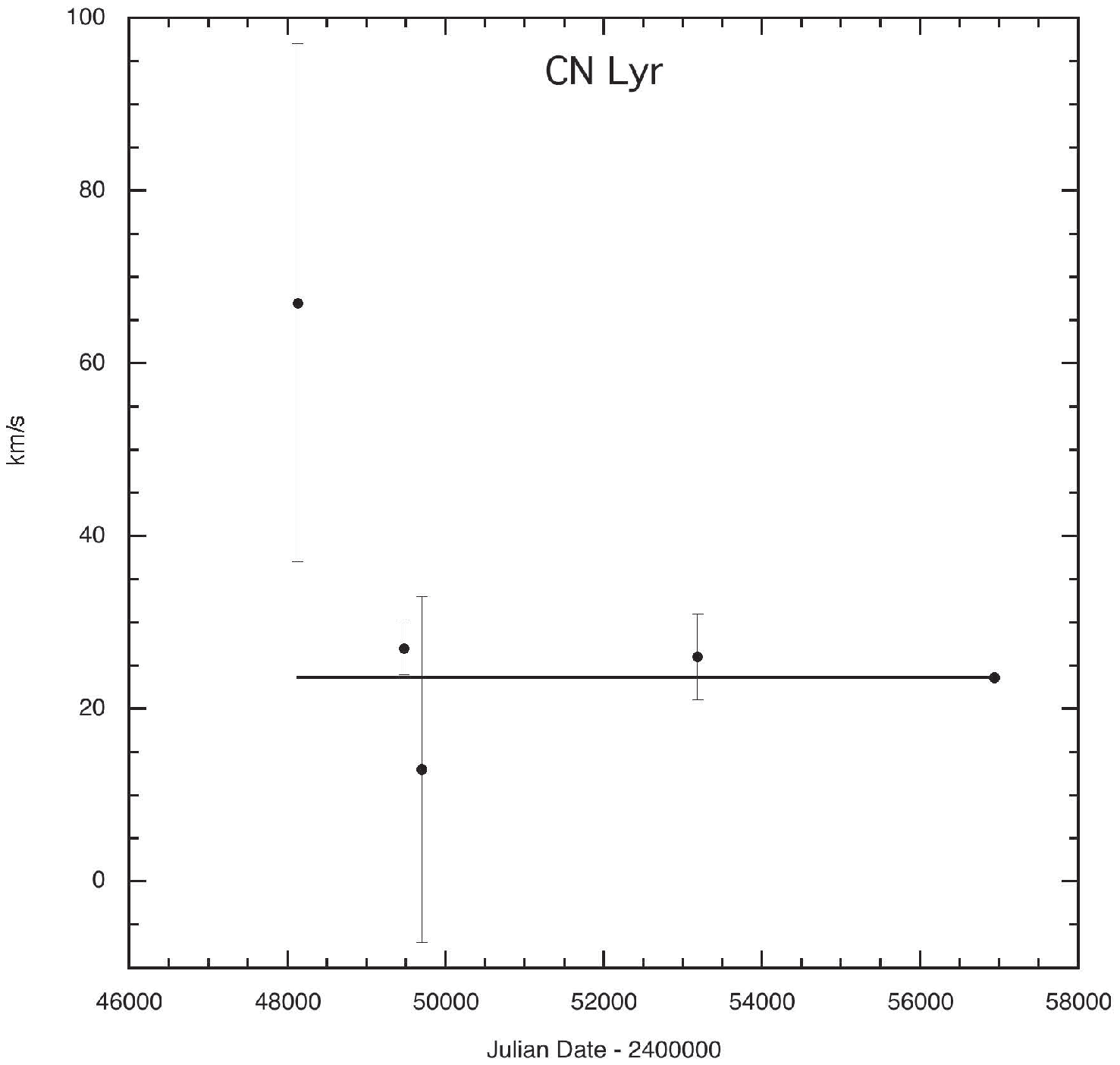}
      \figcaption{Center--of--mass radial velocities for CN Lyr. Some error bars are smaller than the points plotted. The horizontal line shows the weighted mean radial velocity.}
   \label{Fig14}
\end{figure}

\begin{figure}
\figurenum{15}
  \epsscale{0.95}
       \includegraphics[angle=0, width=6.5in]{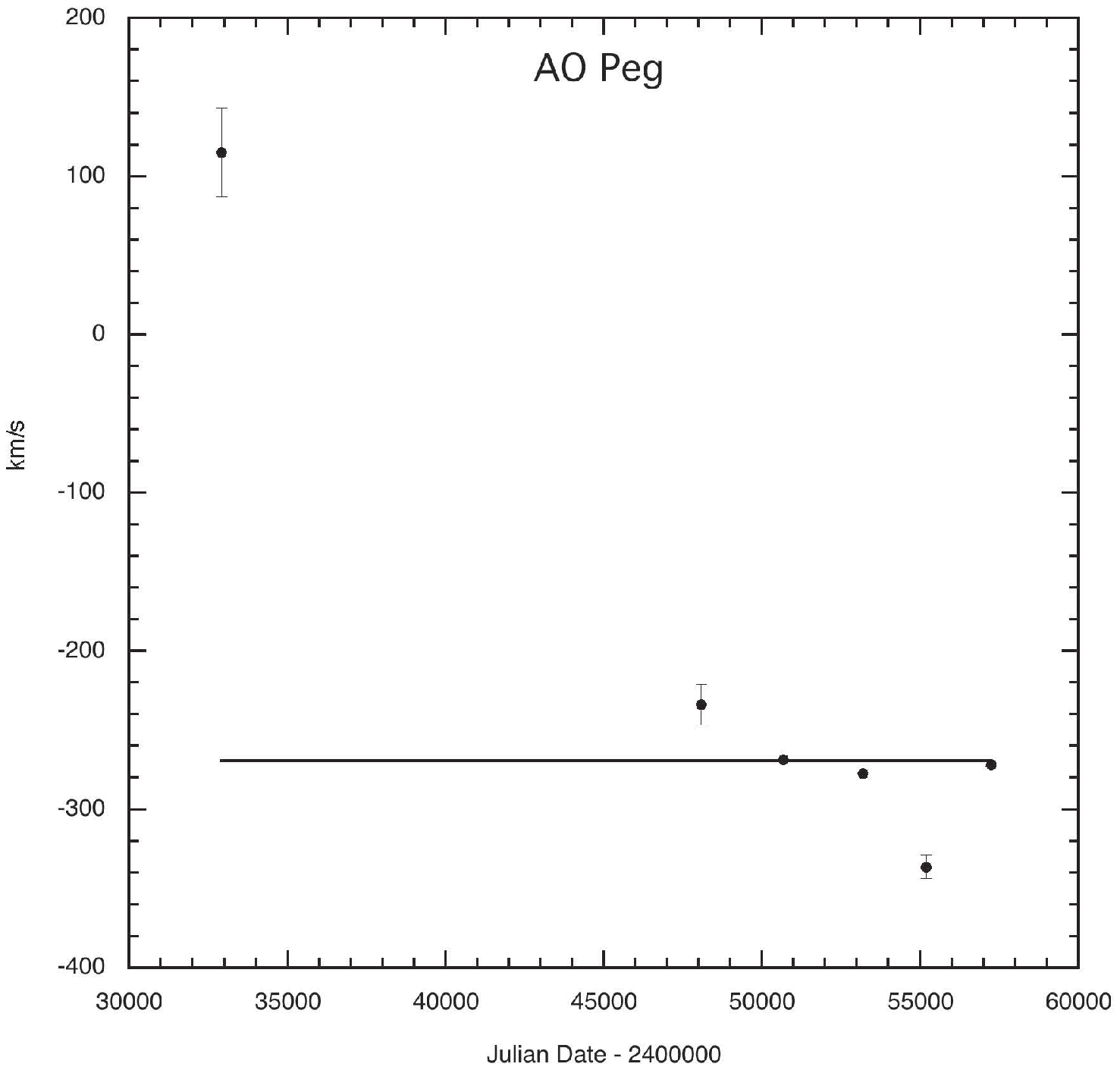}
      \figcaption{Center--of--mass radial velocities for AO Peg. The datum at 55197 for G.--C. Liu {\em et al.} is illustrative only for the JD (see text). Some error bars are smaller than the points plotted. The horizontal line shows the weighted mean radial velocity.}
   \label{Fig15}
\end{figure}

\begin{figure}
\figurenum{16}
  \epsscale{0.95}
    \includegraphics[angle=0, width=6.5in]{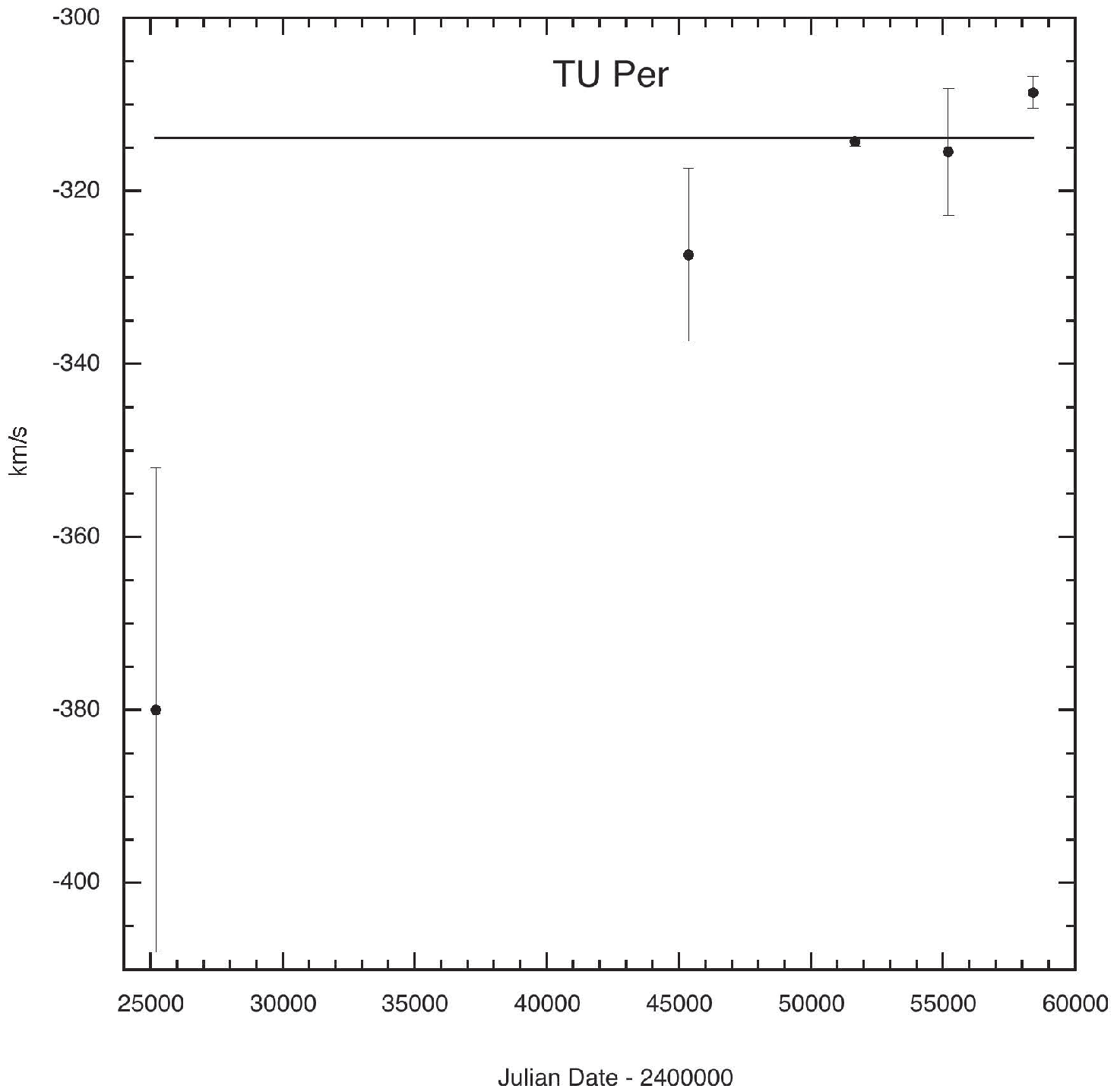}
      \figcaption{Center--of--mass radial velocities for TU Per. The datum at 55197 for G.--C. Liu {\em et al.} is illustrative only for the JD (see text). Some error bars are smaller than the points plotted. The horizontal line shows the weighted mean radial velocity.}
   \label{Fig16}
\end{figure}

\begin{figure}
\figurenum{17}
  \epsscale{0.95}
    \includegraphics[angle=0, width=6.5in]{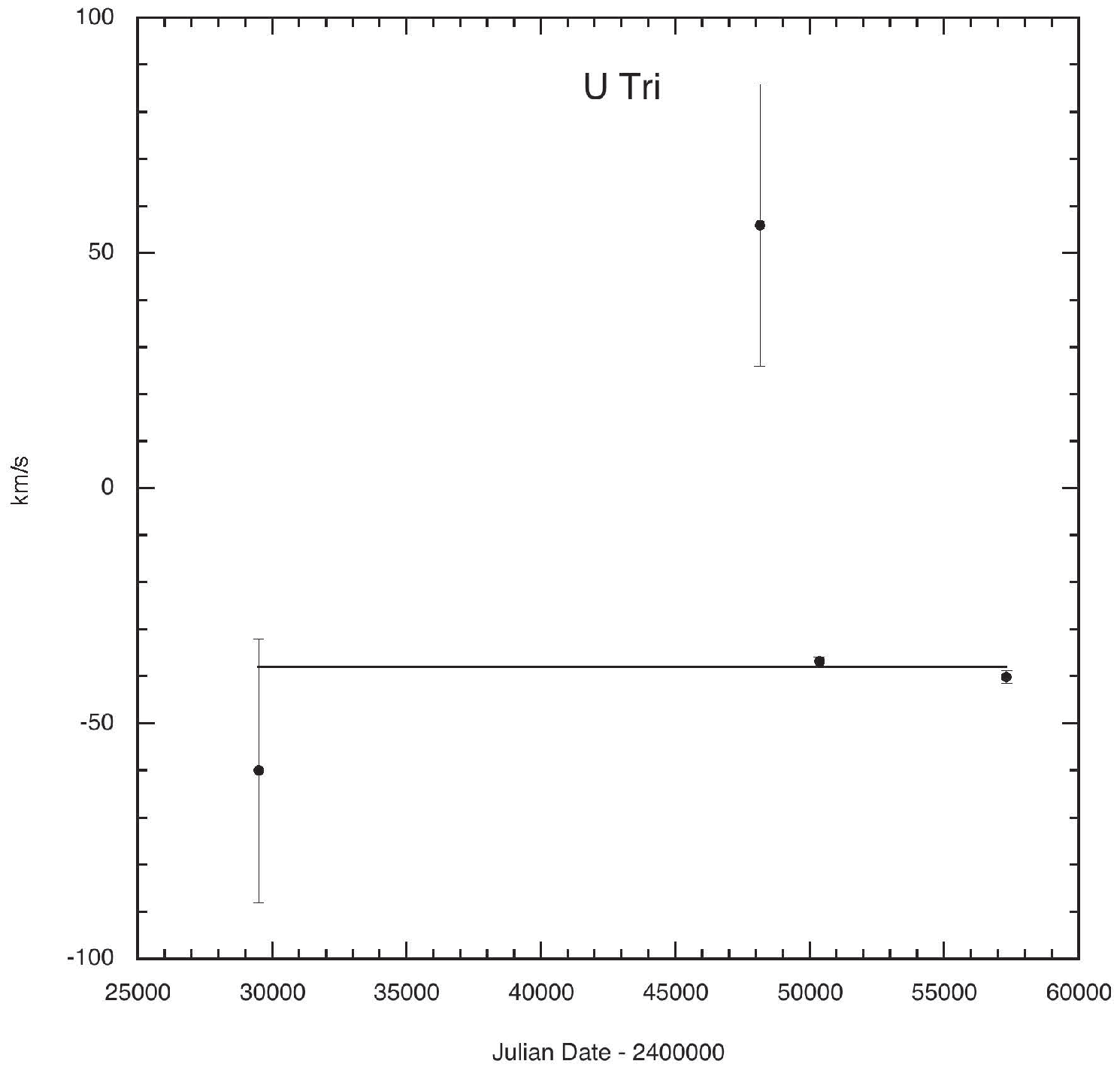}
      \figcaption{Center--of--mass radial velocities for U Tri. Some error bars are smaller than the points plotted. The horizontal line shows the weighted mean radial velocity.}
   \label{Fig17}
\end{figure}

\begin{figure}
\figurenum{18}
  \epsscale{0.95}
    \includegraphics[angle=0, width=6.5in]{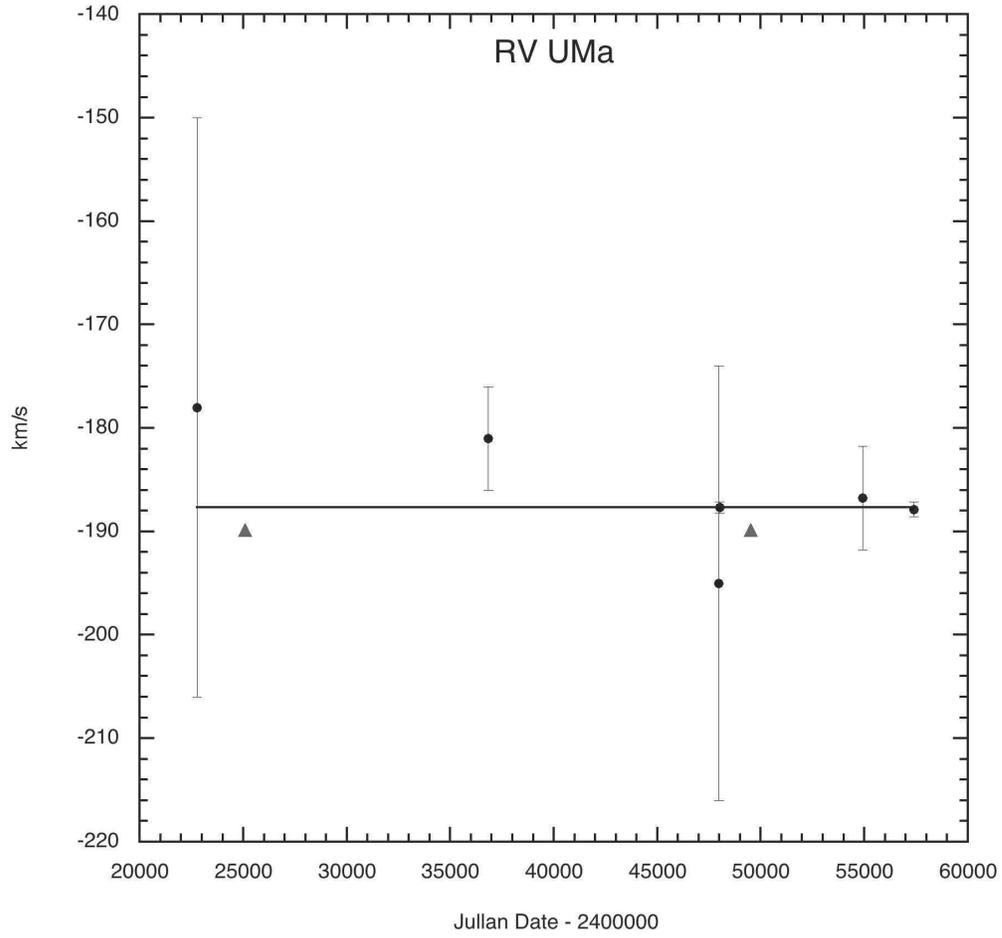}
      \figcaption{Center--of--mass radial velocities for RV UMa. The horizontal line shows the weighted mean radial velocity. The predicted dates of periastron passage and the periastron center--of--mass reflex radial velocity for the 66.9 year orbit are shown as filled triangles.}
   \label{Fig18}
\end{figure}

\begin{figure}
\figurenum{19}
  \epsscale{0.95}
    \includegraphics[angle=0, width=6.5in]{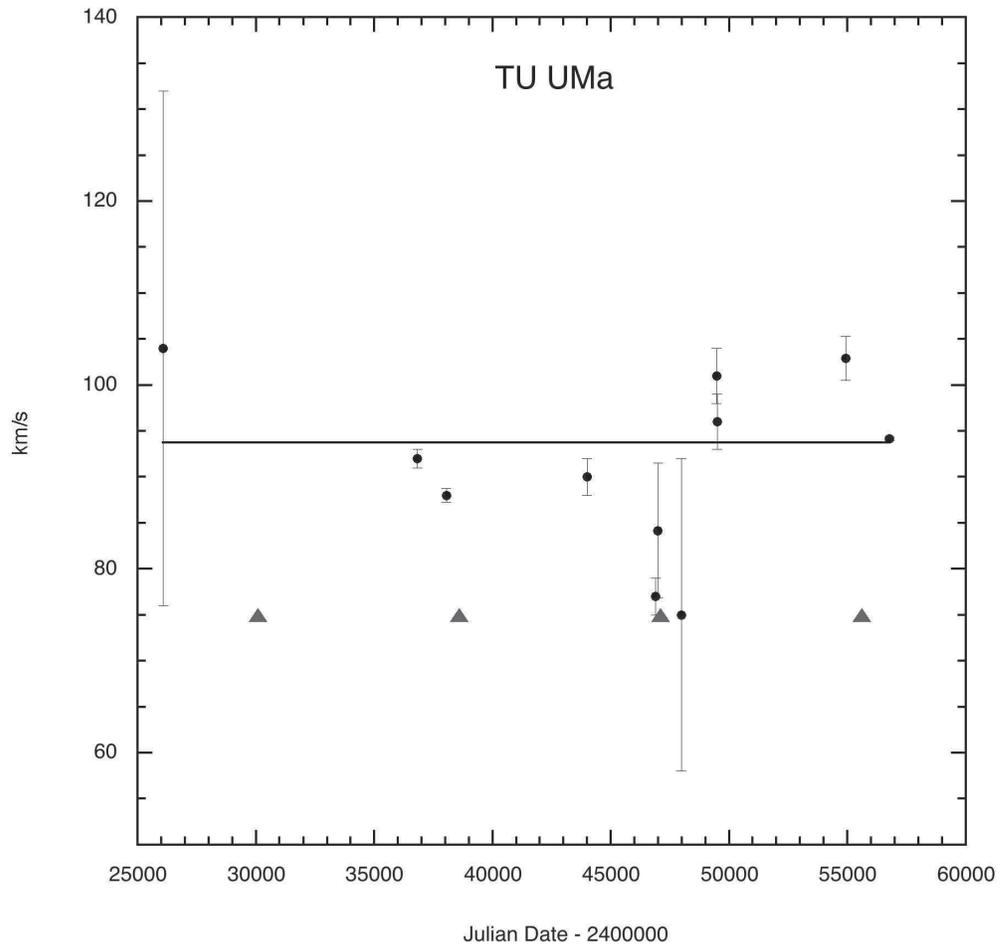}
      \figcaption{Center--of--mass radial velocities for TU UMa. Some error bars are smaller than the points plotted. The horizontal line shows the weighted mean radial velocity. The predicted dates of periastron passage and the periastron center--of--mass reflex radial velocity for the 23.3 year orbit are shown as filled triangles.}
   \label{Fig19}
\end{figure}

\begin{figure}
\figurenum{20}
  \epsscale{0.95}
    \includegraphics[angle=0, width=6.5in]{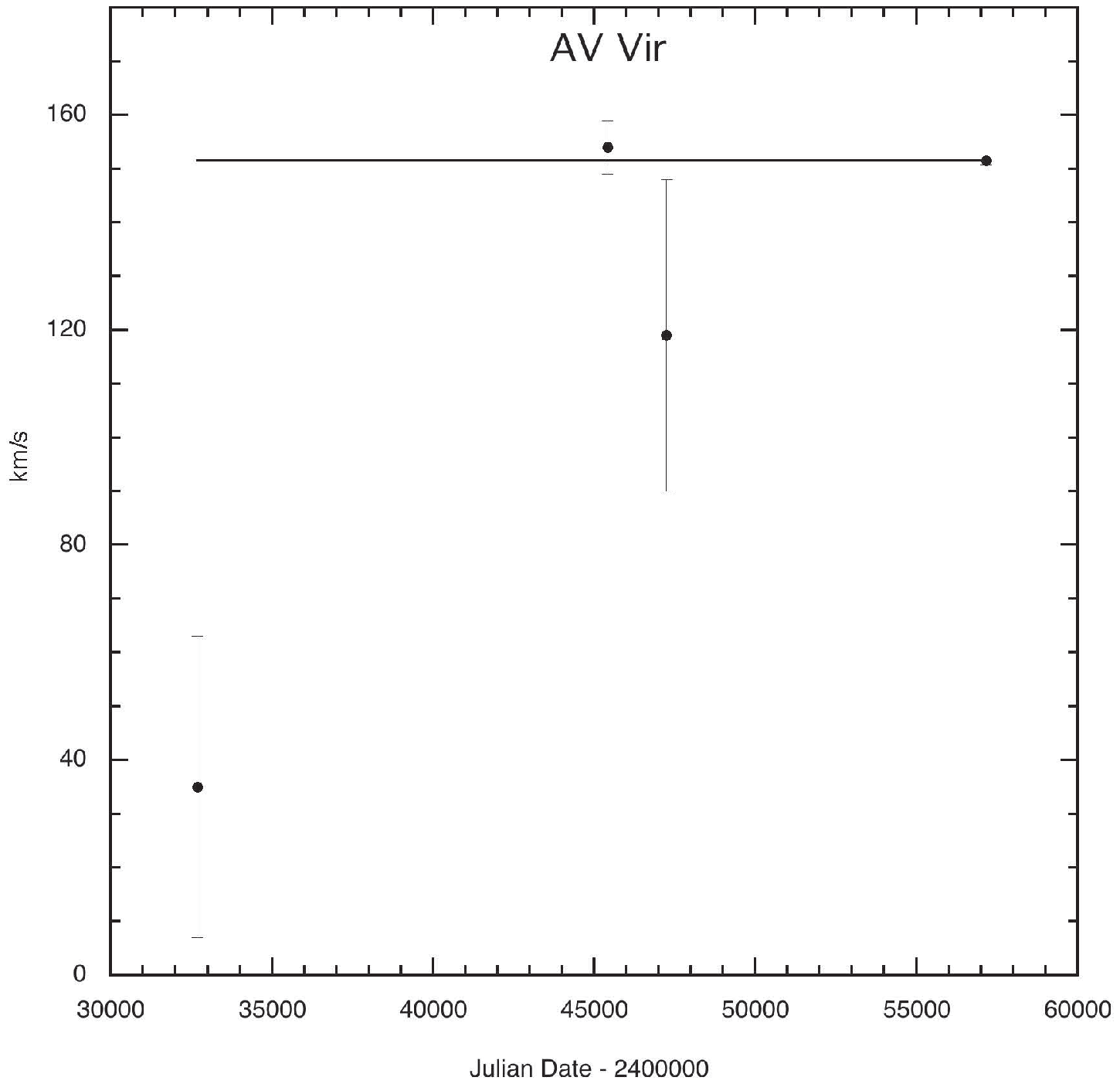}
      \figcaption{Center--of--mass radial velocities for AV Vir. Some error bars are smaller than the points plotted.The horizontal line shows the weighted mean radial velocity.}
   \label{Fig20}
\end{figure}
 
\begin{figure}
\figurenum{21}
  \epsscale{0.95}
    \includegraphics[angle=0, width=6.5in]{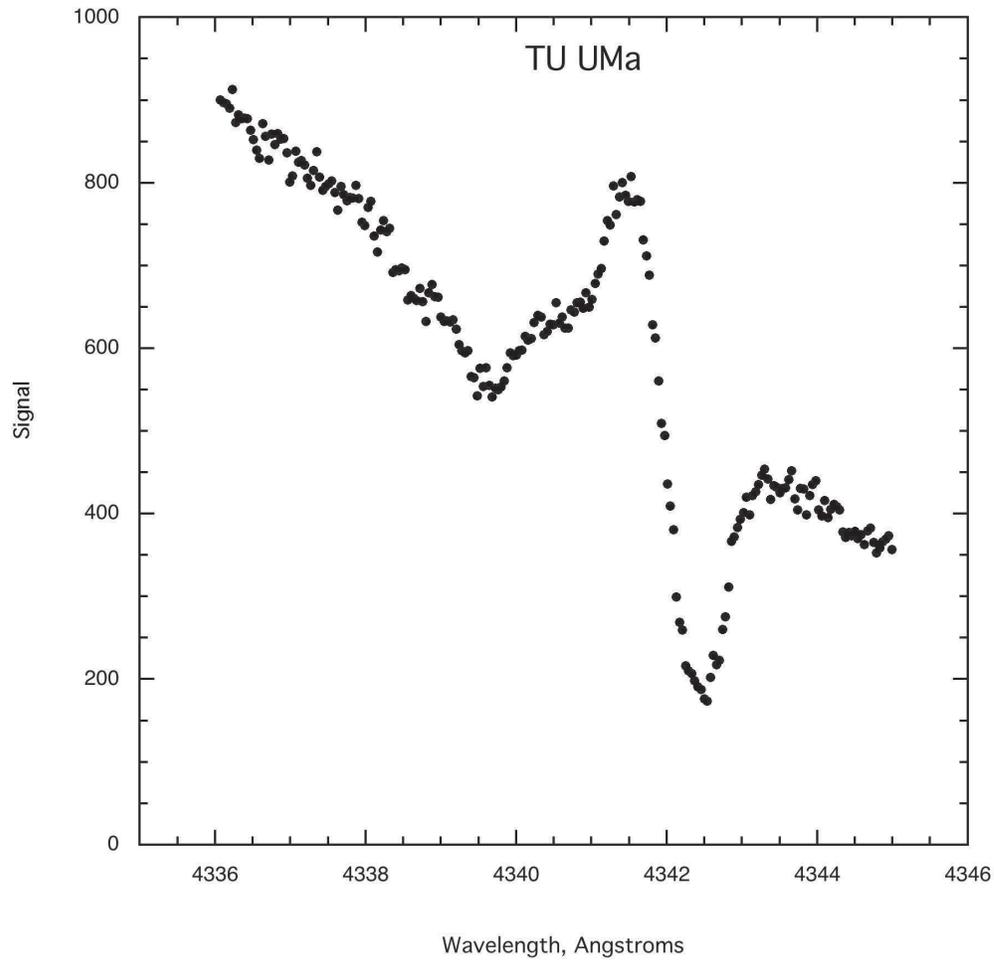}
      \figcaption{$H\gamma$ emission in TU UMa at pulsation phase 0.933. The heliocentric radial velocity for this spectrum is 93.5$\pm 2.9$ km s$^{-1}$  The absorption feature near 4339.5 \AA  is likely Ti II. Instrumental response has not been removed from this spectrum.}
   \label{Fig21}
\end{figure}

\end{document}